\documentstyle[emulateapj]{article}

\font\cap=cmcsc10 at 12pt

\newcommand\hi{{\sc H$\,$i}}
\newcommand\hii{{\sc H$\,$ii}}
\newcommand\MhLb{$M_{HI}/L_B$}
\newcommand\ML{$M_\odot \, L_\odot^{-1}$}
\newcommand\mB{$\mu_B$}
\newcommand\mR{$\mu_R$}
\newcommand\msqas{mag arcsec$^{-2}$}
\newcommand\ch{ch$^{-1}$}
\newcommand\m{$^m$}
\newcommand\kms{km s$^{-1}$}
\newcommand\et{et~al.}
\newcommand\x{$\times$}
\def\+{$\pm$}
\def\col#1{$\times 10^{#1}\, {\rm cm}^{-2}$}
\def\Mo#1{$\times 10^{#1} M_\odot$}

\lefthead{Hibbard, van der Hulst, Barnes \& Rich}
\righthead{{\sc H$\,$i} in The Antennae}
\slugcomment{To Appear in the December 2001 AJ (Vol 122, No. 6)}

\begin{document}

\title{High-resolution {\sc H$\,$i} Mapping of NGC 4038/9 (``The
Antennae") and its Tidal Dwarf Galaxy Candidates}

\author{J.\ E.~Hibbard} 
 \affil{National Radio Astronomy Observatory$^1$,
520 Edgemont Road, Charlottesville, VA, 22903; 
({\it jhibbard@nrao.edu})}
\altaffiltext{1}{The National Radio
Astronomy Observatory is a facility of the National Science Foundation
operated under cooperative agreement by Associated Universities, Inc.}

\and

\author{J.\ M.~van~der~Hulst}
\affil{Kapteyn Astronomical Institute, Postbus 800, NL-9700 AV Groningen,
	The Netherlands; ({\it vdhulst@astro.rug.nl})}

\and

\author{J.\ E.~Barnes} 
\affil{Institute for Astronomy , University of Hawaii,
    Honolulu, HI, 96822; ({\it barnes@ifa.hawaii.edu})}

\and

\author{R.\ M.~Rich} 
\affil{UCLA; ({\it rmr@astro.ucla.edu})}

\begin{abstract}

We present new VLA C+D-array \hi\ observations and optical and NIR
imaging of the well known interacting system NGC 4038/9, ``The
Antennae''.  At low spatial resolution ($\sim 40''$), the radio data
reach a limiting column density of $\sim 10^{19}$ cm$^{-2}\,
(2.5\sigma)$, providing significantly deeper mapping of the tidal
features than afforded by earlier \hi\ maps.  At relatively high
spatial resolution ($\sim 10''$), the radio data reveal a wealth of
gaseous sub-structure both within the main bodies of the galaxies and
along the tidal tails.  In agreement with previous \hi\ studies, we
find that the northern tail has \hi\ along its outer length, but none
along its base.  We suggest that the \hi\ at the base of this tail has
been ionized by massive stars in the disk of NGC 4038.  The gas in the
southern tail has a bifurcated structure, with one filament lying
along the optical tail and another running parallel to it but with no
optical counterpart. The two filaments join just before the location
of several star forming regions near the end of the tail. The \hi\
velocity field at the end of the tail is dominated by strong velocity
gradients which suggest that at this location the tail is bending away
from us.  We delineate and examine two regions within the
tail previously identified as possible sites of a so-called ``tidal
dwarf galaxy'' condensing out of the expanding tidal material. The
tail velocity gradients mask any clear kinematic signature of a
self-gravitating condensation in this region. A dynamical analysis
suggest that there is not enough mass in gas alone for either of these
regions to be self-gravitating. Conversely, if they are bound they
require a significant contribution to their dynamical mass from
evolved stars or dark matter. Even if there are no distinct dynamical
tidal entities, it is clear that there is a unique concentration of
gas, stars and star forming regions within the southern tail: the \hi\
channel maps show clear evidence for a significant condensation near
the tail star forming regions, with the single-channel \hi\ column
densities higher than anywhere else in the system, including within
the main disks. Finally, the data reveal \hi\ emission associated with
the edge-on ``superthin" Scd galaxy ESO 572--G045 that lies just
beyond the southern tidal tail of The Antennae, showing it to be a
companion system. 
\end{abstract}


\section{Introduction}

NGC 4038/9 (= Arp 244), aptly nicknamed ``The Antennae'', is one of
the best-studied examples a galactic collision.  The long tails which
distinguish this system are emblematic of violent tidal interactions
between disk galaxies of similar mass (Toomre \& Toomre 1972,
Schweizer 1978).  Tidal tails originate from the outer regions of
galactic disks and are often rich in neutral atomic hydrogen.  These
features develop kinematically and their velocity fields bear the
imprint of the encounter process, so they provide crucial information
for dynamical modeling of interacting galaxies.  We therefore targeted
NGC~4038/9 for high-spatial and velocity resolution \hi\ observations
(full-width at half maximum, FWHM, as fine as $\sim 10''; \Delta
v\sim$ 5 \kms) in order to obtain the data needed to constrain future
numerical simulations of this system.

NGC 4038/9 was the first major merger to be mapped in the 21-cm line
of neutral hydrogen.  The presence of significant quantities of
neutral hydrogen within both the main bodies and the tails of
NGC~4038/9 was established from single-dish measurements (Peterson \&
Shostak 1974, Huchtmeier \& Bohnenstengel 1975).  Subsequently, NGC
4038/9 has been the subject of \hi\ synthesis mapping observations
using the Westerbork Synthesis Radio Telescope (WSRT; van der Hulst
1979a), the Very Large Array (VLA; Mahoney, Burke \& van der Hulst
1987) and the Australian Telescope Compact Array (ATCA; Gordon,
Koribalski \& Jones 2001).  The observations reported in this paper
provide a more detailed mapping of the tidal features than afforded by
previously published \hi\ maps, representing an improvement in spatial
and velocity resolution by over a factor of two.

Other spectral line studies have mapped the kinematics of NGC 4038/9
using CO spectral lines (Stanford \et\ 1990, Wilson \et\ 2000, Gao
\et\  2001, Zhu 2001) and optical emission lines (Burbidge \& Burbidge
1966, Ruben, Ford \& D'Odorico 1970, Amram \et\ 1992). However, the
molecular and ionized gas is confined to the inner disk regions, and
their kinematics do not constrain the encounter geometry as well as
that of the tidal tails.

At a heliocentric velocity\footnote{Heliocentric velocities are quoted
throughout this paper.} of 1630 \kms\ (corresponding to a velocity
relative to the local group of 1440 \kms), The Antennae is one of the
nearest on-going major mergers.  To be consistent with the majority of
the recent work on NGC 4038/9, we derive physical quantities by
adopting a Hubble Constant of 75 \kms\ Mpc$^{-1}$, yielding a distance
of 19.2 Mpc. However, a recent analysis of $HST$ $VI$ photometry of
individual stars within the southern tidal tail of NGC 4038/9 by
Saviane, Rich \& Hibbard (2001) suggests a much
smaller distance of 13.8\+0.5 Mpc. This distance is
derived by identifying luminous red stars within the tail with the red
giant branch of an old, metal poor population. There are some caveats
to this derivation, for which we refer the reader to Saviane \et, so
for the purposes of the present work, we will use the more commonly
accepted value of 19.2 Mpc.

At this distance, the tidal tails of NGC 4038/9 extend some 65 kpc in
radius and measure $\sim 110$ kpc from end to end.  The bodies of the
galaxies are sites of extensive star formation, producing an IR
luminosity of $\log(L_{\rm IR}) = 10.76$ and an IR to blue luminosity
ratio of $L_{\rm IR}/L_{\rm B} \simeq 6$.  Long-wavelength studies
from the NIR to Radio suggest that this luminosity is powered by an
active system-wide starburst, with no indication of a significant
contribution by an AGN (Hummel \& van der Hulst 1986, Vigroux
\et\  1996, Fischer \et\  1996, Kunze \et\  1996, Mirabel \et\  1998,
Nikola, T. \et\ 1998, Laurent \et\ 2000, Haas \et\ 2000, Xu \et\ 2000,
Neff \& Ulvestad 2000, Mengel \et\ 2001).  While the IR luminosity
is an order of magnitude lower than the most active star forming
mergers (the so-called Ultraluminous Infrared Galaxies), it is still a
factor of $\sim$5 higher than non-interacting galaxy pairs
(e.g.\ Kennicutt \et\ 1987, Bushouse 1987, Bushouse, Lamb \& Werner
1988), with an inferred star formation rate (SFR) of $\sim 20\,
M_\odot \, {\rm yr}^{-1}$ (Evans, Harper \& Helou 1997; Zhang, Fall \&
Whitmore 2001).

NGC 4038/9 is the nearest system with an identified dwarf-galaxy sized
concentration of gas, light and young stars embedded within a luminous
tidal tail (Schweizer 1978, hereafter S78; Mirabel, Dottori, \& Lutz
1992, hereafter MDL92). Such concentrations have been known for
quite some time (e.g.\ Zwicky 1956), but have only recently received
detailed attention. Several numerical studies lend support to the
hypothesis that such systems are self-gravitating and may evolve into
independent dwarf galaxies (Barnes \& Hernquist 1992, Elmegreen,
Kaufmann \& Thomasson 1993), but the supporting observational evidence
is primarily circumstantial.  These observations show that the optical
condensations within tidal tails contain many young stars (S78, MDL92,
Hunsberger, Charlton, \& Zaritsky 1998, Weilbacher \et\ 2000,
Iglesias-P\' aramo \& V\' ilchez 2001, Saviane, Rich \& Hibbard 2001)
and have global properties, such as size, luminosity, \hi\ mass, \hi\
velocity dispersion, and CO content, in common with dwarf Irregular
galaxies (Mirabel \et\ 1992; Hibbard \et\ 1994; Smith \& Higdon 1994;
Duc \& Mirabel 1994, Duc \et\ 1997, 2000; Smith \& Struck 2001; Braine
\et\ 2000, 2001).  Recently, some tidal dwarfs have even been found to
harbor ``Super Star Clusters'' (Knierman \et\ 2001). For these
reasons, such concentrations are often referred to as ``Tidal Dwarf
Galaxies'' (hereafter TDGs).  However, strong kinematic evidence that
these concentrations are indeed self-gravitating is lacking, and their
evolution into dwarfs is therefore still in question (see Hibbard,
Barnes, \& van der Hulst 2001, hereafter Paper II, for more
details). Another objective of the present observations is therefore
to search for the kinematic signature of a self-gravitating mass in
the vicinity of the putative TDG.

The paper is organized in the following manner.  In \S2 we present
details on the observations and data reduction procedures.  In \S3 we
describe in turn the morphology and kinematics of the tidal tails,
sub-structure within the tails, the region in the vicinity of the TDG,
the inner disks, and the companion galaxy. In \S4 we try to explain
various aspects of the \hi\ tidal morphology, and look for evidence
for a kinematically distinct entity at the end of the southern
tail. Our conclusions are presented in \S5.

\section{Observations}

\subsection{VLA \hi\  Observations}

The Antennae was observed in two separate array configuration of the
VLA radio interferometer: the most compact configuration (1 km
D-array) in June of 1996, and the hybrid CnB-array in June of
1997. For the hybrid array, the antennae along the E and W arms of the
array are at the standard C-array (3 km) stations while those of the N
arm are at the more extended B-array (9 km) stations. This compensates
for the foreshortening of baselines to the north arm antennae that
occurs for sources at low declination.  The short baselines of the
D-array provide the highest sensitivity to extended gas, while the
longer baselines of the CnB-array allow resolutions as fine as
$\sim10''$.  NGC 4038/9 was observed for a total of 3.5 hours in the
D-array and 8 hours in the CnB-array.  The details of these
observations are tabulated in Table~\ref{tab:HIobs}.

We observed using the 21-cm spectral line mode with the 
system tuned to a central frequency corresponding to a
heliocentric velocity of 1630 \kms, the central velocity of the
existing \hi\  synthesis observations (Mahoney \et\  1987). Since the
atomic gas within tidal tails typically has narrow linewidths
($\sigma_v \sim$ 5---10 \kms; Hibbard \et\  1994, Hibbard \& van
Gorkom 1996), the correlator mode was chosen to give the smallest
channel spacing while still covering the velocity spread of existing
\hi\  observations (1360--1860 \kms; van der Hulst 1979a, Mahoney \et\  1987).
New broader-bandwidth \hi\ (Gordon \et\ 2001) and CO
(Gao \et\ 2001) observations reveal that there is both atomic and
molecular gas over the velocity range from 1340--1945 \kms. We discuss
the implications of this result below.

Using on-line Hanning smoothing, the chosen set-up provides 127
independent velocity channels over the 3.125 MHz bandwidth, resulting
in a channel spacing of 5.21 \kms. The pointing center was chosen to
place the ends of the optical tails at equal distances from the phase
center.  This position is $2\farcm5$ south of the main bodies, and
places the ends of the optical tails at a radius of $\sim 10'$,
which is approximately the 75\% sensitivity point of the primary beam
(see Figure \ref{fig:HIdss}).

The data were calibrated, mapped, and ``Cleaned" using standard
techniques and procedures in the Astronomical Image Processing System
(AIPS) (see e.g.~Rupen 1999).  The data from the separate arrays were
calibrated and bandpass corrected independently, after which they were
combined in the UV plane to produce the C+D dataset. This latter
dataset is used exclusively throughout this paper. The final
calibration of this dataset was achieved by three iterations of a
phase-only self-calibration. The resulting continuum map is very
similar to others that have appeared in the literature (e.g. Hummel \&
van der Hulst 1986; Neff \& Ulvestad 2000, Gordon \et\ 2001), and is
not shown here.

The continuum was subtracted from the velocity cube by fitting a
first-order polynomial to the visibilities from the line free channels
on either end of the bandpass.  This is done in an iterative manner in
order to determine which channels are free of line emission. The
resulting fit was made using channels 11--20 and 104--111 of the 127
channel cube, corresponding to the velocities 1859--1911 \kms\ and
1386--1423 \kms. As mentioned above, cold gas has been found within
both of these velocity ranges by Gordon \et\ (2001) and Gao \et\
(2001). The gas at these velocities is of low surface brightness and
totally confined to either the disk of NGC 4039 or the disk overlap
region. Therefore, using this continuum range will not affect any of
the measured properties of the tidal tails, which is our primary
interest in this paper.  However, the \hi\ flux we measure for the
central region will underestimate the true flux, and the kinematics of
this region will not be accurately mapped (although the fact that the
line kinematics are weighted by intensity will somewhat mitigate the 
effects of the missing velocity information).

The continuum subtracted data were mapped using several values of the
Robust weighting parameter, $R$ (Briggs 1995).  This parameter can be
varied to emphasize either the outer ($R<0$) or the inner ($R>0$)
regions of the UV plane (i.e., either small or large spatial scales).
In practical terms, a lower $R$ gives a finer spatial resolution at
the expense of surface brightness sensitivity, while a larger $R$
gives a higher surface brightness sensitivity at a slightly coarser
spatial resolution.  A high resolution data cube was made with $R=-1$,
providing a surface brightness rms sensitivity of 1.3 mJy beam$^{-1}$
at a resolution of $\theta_{FWHM}=11.4''\times 7.4''$.  This
corresponds to a single channel column density limit (2.5$\sigma$ per
channel) of 2.2\col{20} averaged over the beam width of 1.1\x0.7
kpc$^2$.  A more sensitive intermediate resolution data cube, made
with $R=+1$, provides a resolution of $21''\times 15''$ (1.9\x 1.4
kpc$^2$) and a column density limit of 4\col{19}.  To further increase
sensitivity to extended low column density gas, a low resolution data
cube was made by convolving the $R=+1$ data cube to a resolution of
$40''$ (3.7 kpc), reaching a detection limit of 1.2\col{19}.  These
data will be referred to in the following as the high, intermediate,
and low resolution data, respectively.  These parameters are
summarized in Table~\ref{tab:HIobs}.  The resulting per channel noise
is close to theoretical once the higher system temperature due to the
low declination of the source is taken into consideration.

Visualization of the resulting data cubes and inter-comparison with the
optical and NIR data was greatly facilitated by using the {\sc Karma}
visualization package (Gooch 1995).  Besides allowing multiple
images to be interactively compared, this package also allows a full
3-dimensional rendering of datacubes, which is a particularly powerful
tool for disturbed systems.

After a careful viewing, the data cubes were integrated
over the velocity axis to produce the moment maps.  In order to
suppress the effects of noise, only data above a fixed threshold are
used in the moment summation (the ``cutoff technique'' described by
Bosma, 1981) using the AIPS task MOMNT. This task applies a user
specified threshold to a version of the data cube smoothed in both
space and velocity.  If a pixel in the smoothed data cube passes the
threshold, the corresponding pixel in the un-smoothed data is included
in the moment analysis.  This procedure favors low level emission that
is extended in velocity and/or space over emission which is
isolated. The output from this analysis is an integrated intensity map
(zeroth moment), an intensity-weighted velocity map (first moment),
and an intensity-weighted velocity dispersion map (second moment). It
should be noted that the first and second moment maps give an accurate
representation of the mean \hi\  velocity and line-of-sight velocity
dispersion at a location only if the line profiles are single-peaked.
The final zeroth moment maps were corrected for
the primary beam attenuation before measuring \hi\  fluxes.

We measure \hi\ fluxes from an integrated intensity map constructed to
match the new ATCA \hi\ observations of Gordon \et\ (2001), since
those authors were able to do a proper line-free continuum
subtraction.  For the moment analysis, this involved applying a
threshold of 2.1 mJy beam$^{-1}$ to the data after smoothing with a
3\x3 pixel boxcar spatial filter and a five-channel Hanning filter.
Our resulting flux measurements are given in Table~\ref{tab:global}.
The errors attached to the flux measurements have been determined from
the single channel noise by properly taking into account the number of
independent channels and beams over which the flux was integrated. It
should be noted that the error calculation does not include systematic
errors, such as those introduced by poor continuum subtraction or
incomplete cleaning.

\newpage
\subsection{Optical and NIR Observations}

The supporting optical observations were obtained during two nights of
broad-band $BVR$ imaging on the CTIO 0.9m with the Tek512 CCD in April
of 1991 and one night of near-Infrared (NIR) $K'$-band imaging on the
UH $88''$ with the QUIRC detector in January of 1995. The details of
these runs are given in Table~\ref{tab:OPTobs}. For the optical run
the conditions were photometric and the seeing was $\sim 1''$.  The
f/13.5 re-imaging optics were used, giving a plate scale of $0.44''$
pixel$^{-1}$ and a field of view of 3.8$'$.  The data were calibrated
via observations of standards in the Graham E-regions (Graham 1982)
observed on the same nights, with zeropoint errors (1$\sigma$) of
0.04\m\ in $B$ and 0.02\m\ in both $V$ and $R$. For the NIR run, the
data were obtained through a $K'$ filter ($\lambda = 2.11 \mu$m,
$\Delta \lambda = 0.35 \mu$m; Wainscoat \& Cowie 1992; hereafter
referred to simply as $K$).  The f/10 re-imaging optics were used,
resulting in a plate scale of $0.187''$ pixel$^{-1}$ and a field of
view of 3\farcm2. The seeing was $\sim 1''$. These observations
consist of three 120 sec target-sky pairs, with the CCD dithered by
1\farcm5 between on-source positions. The NIR data are uncalibrated.

The combined images were transformed to the World Coordinate System
(and thereby to the same reference frame as the radio images) by
registering to an image of the same area extracted from the Digitized
Sky Survey\footnote{The Palomar Observatory Sky Survey was made by the
California Institute of Technology with funds from the National
Science Foundation, the National Geographic Society, the Sloan
Foundation, the Samuel Oschin Foundation, and the Eastman Kodak
Corporation.  The Oschin Schmidt Telescope is operated by the
California Institute of Technology and Palomar Observatory.}~(DSS).
The registration was accomplished by referencing the location of
$\sim$ 20 stars in common on both images using the {\sc koords} program
in the {\sc Karma} package (Gooch 1995) to perform a non-linear least
squares fit for the transformation equations.  The plate solution so
found is accurate to a fraction of a pixel, and the overall
registration should be as good as that of the southern portion of the
Guide Star Catalog, which is estimated to be $0\farcs7$ (Taff \et\
1990).

Deep optical images were constructed by applying a 9\x9 pixel boxcar
median to the pixels with the lowest light levels, achieving limiting
surface brightnesses of \mB=27.0 \msqas\  and \mR=26.5 \msqas.  Color
maps in $B-R$ and $B-K$ were made after convolving the images to a
common resolution, and using only pixels with a signal-to-noise
greater than five in both bands.  These maps will be shown here, but a
full color analysis is deferred to a later paper (Evans \et, in
prep.).  Finally, a smoothed $B$-band image was made by replacing
stars with background values and convolving to the resolution of the
intermediate resolution \hi\  data. This image is used to evaluate the
\hi\  mass to blue light ratio ($M_{HI}/L_B$) of various regions.

\section{Results}

\begin{figure*}[t!]
\plotone{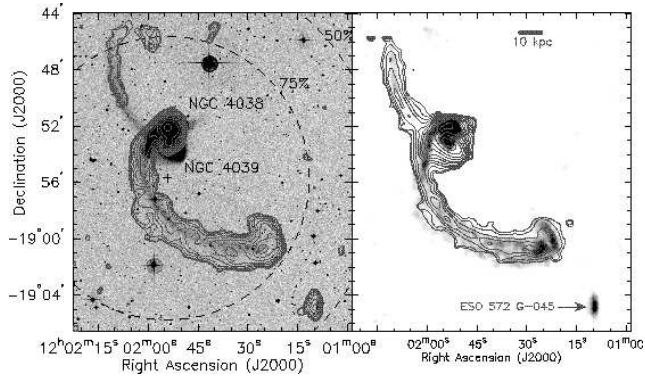}
\figcaption{Left: Integrated \hi\  emission contoured upon
on a greyscale representation of the DSS image of ``The Antennae''.
The cross marks the phase center of the VLA observations, while the
dashed circles represent the 75\% and 50\% response points of the VLA
primary beam. The two main disks, NGC 4038 and NGC 4039, are labeled,
as is the companion galaxy ESO 572 G--045. The contours are from the 
low resolution (40$''$) datacube, and are are drawn at flux
levels of (5, 10, 20, ...) \x 14.49 mJy beam$^{-1}$
\kms, with the lowest contour corresponding to an \hi\ column density
of 5\col{19}. Right: Greyscale representation of the intermediate
resolution \hi\ ($20.7''\times 15.4''$), with contours from the
$B$-band image, after star subtraction and convolving to a resolution
of $25''$. Optical contours are drawn from 27 \msqas\ to 20.5 \msqas\
in intervals of 0.5 \msqas. The box-like optical contours to the
northwest of the disk indicate the extent of the CCD frame in this
direction. The scale bar indicates a length of 10 kpc for a distance of
19.2 Mpc.
\label{fig:HIdss}}
\end{figure*}

The large scale distribution of the \hi\  is shown in Figure
\ref{fig:HIdss}, where the left panel shows the low resolution 
\hi\  contoured upon the DSS image, while the right panel shows the
intermediate resolution \hi\ in greyscales with contours from the
smoothed $B$-band superimposed. Our new observations delineate the
distribution of tidal \hi\ much more clearly than earlier \hi\
observations, showing that the gas in the northern tail extends
further than previously known, and revealing a small \hi-rich disk
companion to The Antennae lying just beyond the southern tail. These
results are confirmed by the ATCA observations by Gordon \et\ (2001).
Using the NASA Extragalactic Database (NED), this latter object is
identified with the galaxy ESO 572--G045.  We break the \hi\ emission
into four separate components: the southern tail, the
northern tail, the inner disks, and the new dwarf companion ESO
572--G045. The \hi\ properties of each of these components are listed
in Table~\ref{tab:global}.

\begin{figure*}
\plotfiddle{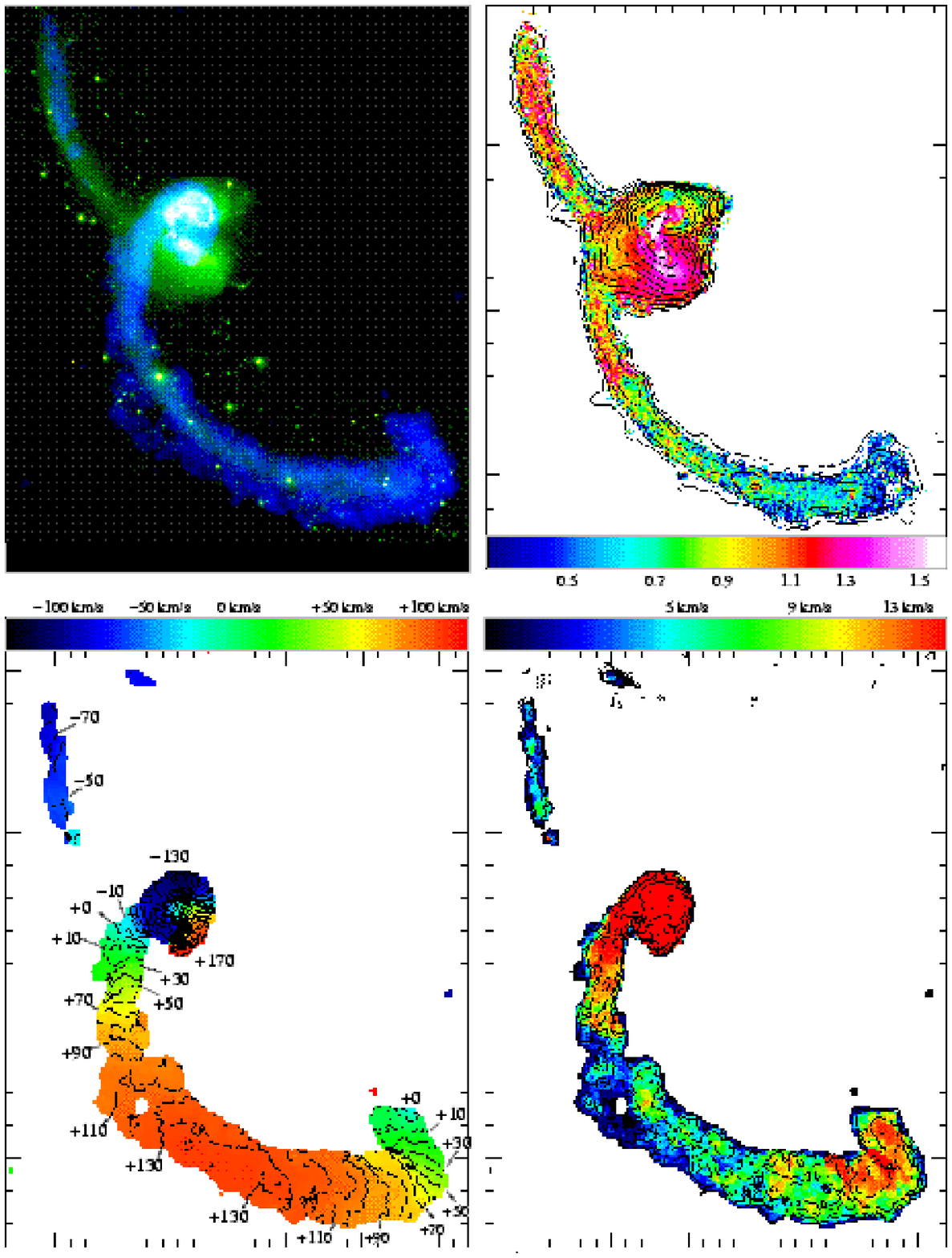}{20cm}{0}{97}{97}{-290}{-88}
\figcaption{Large scale distribution of the \hi\  and optical light in 
NGC 4038/9.  The upper left panel shows a false color representation
of the intermediate resolution \hi\ data (blue), and the CTIO $B+V+R$
combined image in green and white.  The upper right panel of shows the
$B-R$ color map, where colors have been determined for pixels with a
S/N $>$ 5 ($\mu_B <$ 26.4 \msqas). Blue colors correspond to $[B-R <
0.5]$, light blue/cyan to $[0.5 < B-R < 0.7]$, green to $[0.7 < B-R <
0.9]$, yellow/orange to $[0.9 < B-R < 1.1]$, red to $[1.1 < B-R <
1.3]$, magenta to $[1.3 < B-R < 1.5]$, and white to $[B-R >
1.5]$. Surface brightness contours from the star-subtracted $B$-band
image convolved to 25$''$ resolution are drawn at intervals of (26.5,
26, 25.5, ..., 20.5) \msqas.  The lower left panel shows the intensity
weighted \hi\ velocities (green indicates velocities close to
systemic, and red and blue indicate redshifted and blueshifted
velocities, respectively) with isovelocity
contours drawn at 10 \kms\ intervals. The lower right panel maps the
\hi\ velocity dispersion with the intermediate \hi\ column density
contours superimposed. In this panel blue/cyan corresponds to
[$\sigma_{HI} < 5.0$ \kms], green to [$5.0 < \sigma_{HI} < 9.0$ \kms],
yellow/orange to [$9.0 < \sigma_{HI} < 13.0$ \kms], and red to
[$\sigma_{HI} > 13.0$ \kms]. \hi\ column densities are drawn at 
levels of (1, 2, 4, 8, ...) \x 28.88 mJy beam$^{-1}$ \kms, where the lowest
contour corresponds to 1\col{20}.
\label{fig:ALLmos}}
\end{figure*}

The large scale distribution and kinematics of the \hi\  derived from
the intermediate resolution data are shown in Figure \ref{fig:ALLmos}.
The upper left panel shows a false color representation of the data,
with the \hi\  shown in blue and starlight in green and white.  The
upper right panel of Fig.~\ref{fig:ALLmos} shows the $B-R$ color map
with $B$-band surface brightness contours superimposed. The lower left
panel shows a color representation of the intensity weighted \hi\
velocity field with iso-velocity contours superimposed.  Finally, the
lower right panel shows a color representation of the \hi\  velocity 
dispersion, again with \hi\  column density contours superimposed.

As revealed by the WSRT observations of van der Hulst (1979a), the
majority of the \hi\ is associated with NGC 4038 and the southern
tail, with much less gas directly associated with NGC 4039 and the
northern tail. We detect tidal \hi\ emission over a velocity range of
1420 -- 1850 \kms, similar to that found in the synthesis observations
of Mahoney \et\ (1987), but less than found in the broader bandwidth
observations of Gordon \et\ (2001). We measure a total \hi\ flux of at
least 54.5\+0.5 Jy \kms\ for NGC 4038/9, corresponding to a \hi\ mass
of $>$4.7\Mo{9} (Table~\ref{tab:global}).  This is not very different
from the total flux measurement of 57.1 Jy \kms\ measured at the ATCA
by Gordon \et\ (2001). However, when we divide the emission among the
different components we do find differences. We measure a total flux
of 49.7 Jy \kms\ associated with the main disks and southern tail,
compared to the 55.1 Jy \kms\ measured by Gordon \et. Examining the
channel maps of Gordon \et, we find a mean flux level of $\sim$ 10 mJy
beam$^{-1}$ (for a 40$''$ beam) over the velocity range which we used
for continuum subtraction. This suggest that, due to our improper
continuum subtraction, we miss a total of 6 Jy \kms\ over the entire
600 \kms\ velocity range mapped by Gordon \et, which explains the
different disk flux measurements. We also find differences for the
northern tail and companion galaxy, where we recover over twice the
flux found by Gordon \et.

We next present details of the observations for each of the components
individually, as well as for the tidal dwarf candidate(s)
within the southern tail.

\subsection{Southern Tail Morphology}
\label{sec:Stail}

The majority of the the tidal gas is associated with the southern
tail.  This tail extends to a projected radius of $11\farcm6$ (65
kpc), and contains 2.8\Mo{9} of \hi\ along its entire length. The
column density slowly decreases with increasing distance along the
tail, reaches a minimum, and increases again near its end.
Perpendicular to the tail, the gas is distributed more broadly than
the optical light, with a notable extension running along the outer
(southern) edge of the tail.  There is no optical counterpart to this
outer extension at our limiting surface brightness of $\mu_B <$ 27
\msqas\ (see Fig.~\ref{fig:HIdss}), with a resulting \hi\ gas-to-light
ratio of \MhLb $>$ 1.5 \ML, compared to $\sim$0.5 \ML\ on the optical
tail.

\begin{figure*}
\plotone{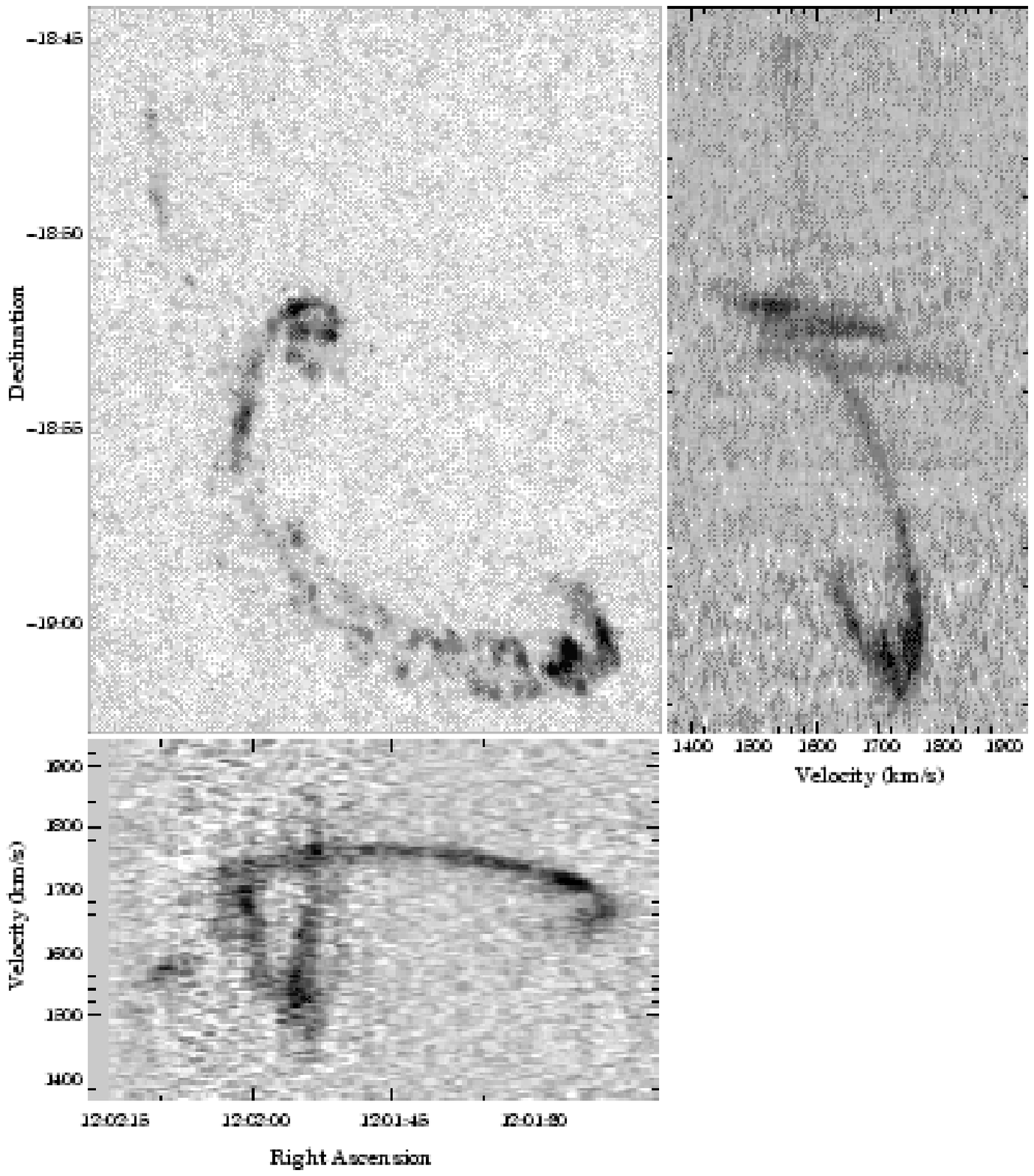}
\figcaption{
Three orthogonal projections of the the intermediate resolution \hi\
datacube. The upper left panel shows the sky view, $I(\alpha,
\delta)$, and to either side of this are the two orthogonal
position-velocity profiles, $I(\alpha, V_z)$ (bottom) and $I(V_z,
\delta)$ (right). The $(\alpha, \delta)$ view is displayed using a
``maximum voxel'' function, which displays the value of the brightest
pixel along the third ($V_z$) dimension (see text). The $I(\alpha,
V_z)$ and $I(V_z, \delta)$ projections are constructed by summing the
emission along one of the spatial dimensions of the datacube, either
right ascension or declination. These panels clearly show the 
well-behaved kinematic structure of the tidal features, as well as the
wealth of gaseous sub-structure within the tails and disk.
\label{fig:posvel}}
\end{figure*}

The gas within the southern tail shows considerable structure on the
scale of one beamwidth ($\sim$ 1--2 kpc). This is further illustrated in
Figure \ref{fig:posvel}, which displays three orthogonal projections of
the intermediate resolution \hi\  datacube. The upper left panel 
shows the sky view, $I(\alpha, \delta)$, and to either side of this
are the two orthogonal position-velocity profiles, $I(\alpha, V_z)$
(bottom) and $I(V_z, \delta)$ (right). The $(\alpha, \delta)$ view is
displayed using a ``maximum voxel'' function. This function assigns to
each pixel the maximum intensity found along the velocity
axis. The position-velocity plots are constructed in the more
traditional manner, summing the emission along one of the spatial
dimensions of the datacube, either right ascension or declination. We
will use these plots below when examining the tail kinematics (\S
\ref{sec:TailKin})

The maximum voxel display emphasizes regions where the \hi\  is cold
and dense (maximum number of \hi\  atoms per unit velocity).  As a
result, it emphasizes the dynamically cold tidal tails.  Note
especially how the disk \hi, which has the highest integrated \hi\
column density, appears fainter in Fig.~\ref{fig:posvel} than in the
zeroth moment maps (Fig.~\ref{fig:HIdss}). This is because the disk
gas is spread over a very broad range of velocities.
Fig.~\ref{fig:posvel} reveals a number of dense knots within the
southern tail, especially toward its end.  The maximum voxel display
ensures that this is not due to a line-of-sight integration effect
where the tail bends back along our line-of-sight.  Notice also that
similar dense knots are not found within the northern tail.  The
southern tail also exhibits an interesting parallel or ``bifurcated''
structure that starts where the southern tail begins to bend westward,
and joins back together just before the location of the 
star forming regions identified by Mirabel \et\ (see 
Fig.~\ref{fig:TDGmos} for the precise location of the star forming 
regions). 

\begin{figure*}
\plotone{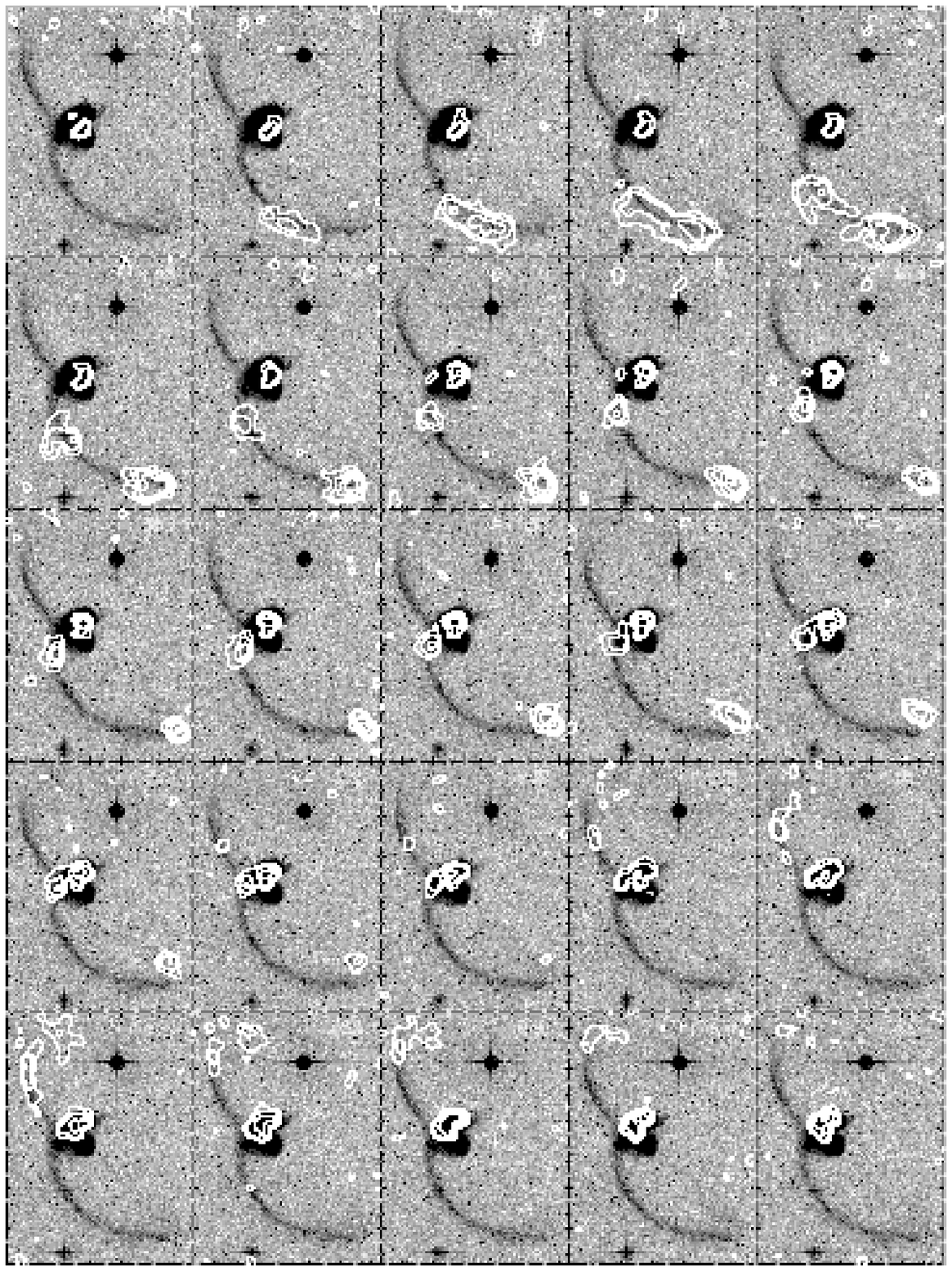}
\figcaption{
\hi\  channel maps of the tidal regions of NGC 4038/9
contoured upon the blue DSS image.  The more sensitive low resolution
\hi\ cube is used after smoothing by a factor of two in velocity (to a
10.6 \kms\ channel width).  The 40$''$ beam size is indicated in the
lower left hand corner of the first panel, and each panel is labeled
with its heliocentric velocity.  Contours are drawn at levels of (3,
6, 12, 24, ...) \x 1.1 mJy beam$^{-1}$, where 1.1 mJy beam$^{-1}$ is
the single channel noise level, corresponding to a column density of
8\col{18}.  Only channels showing {\hi} emission from the tails are
shown.
\label{fig:ALLchan}}
\end{figure*}

\begin{figure*}
\plotone{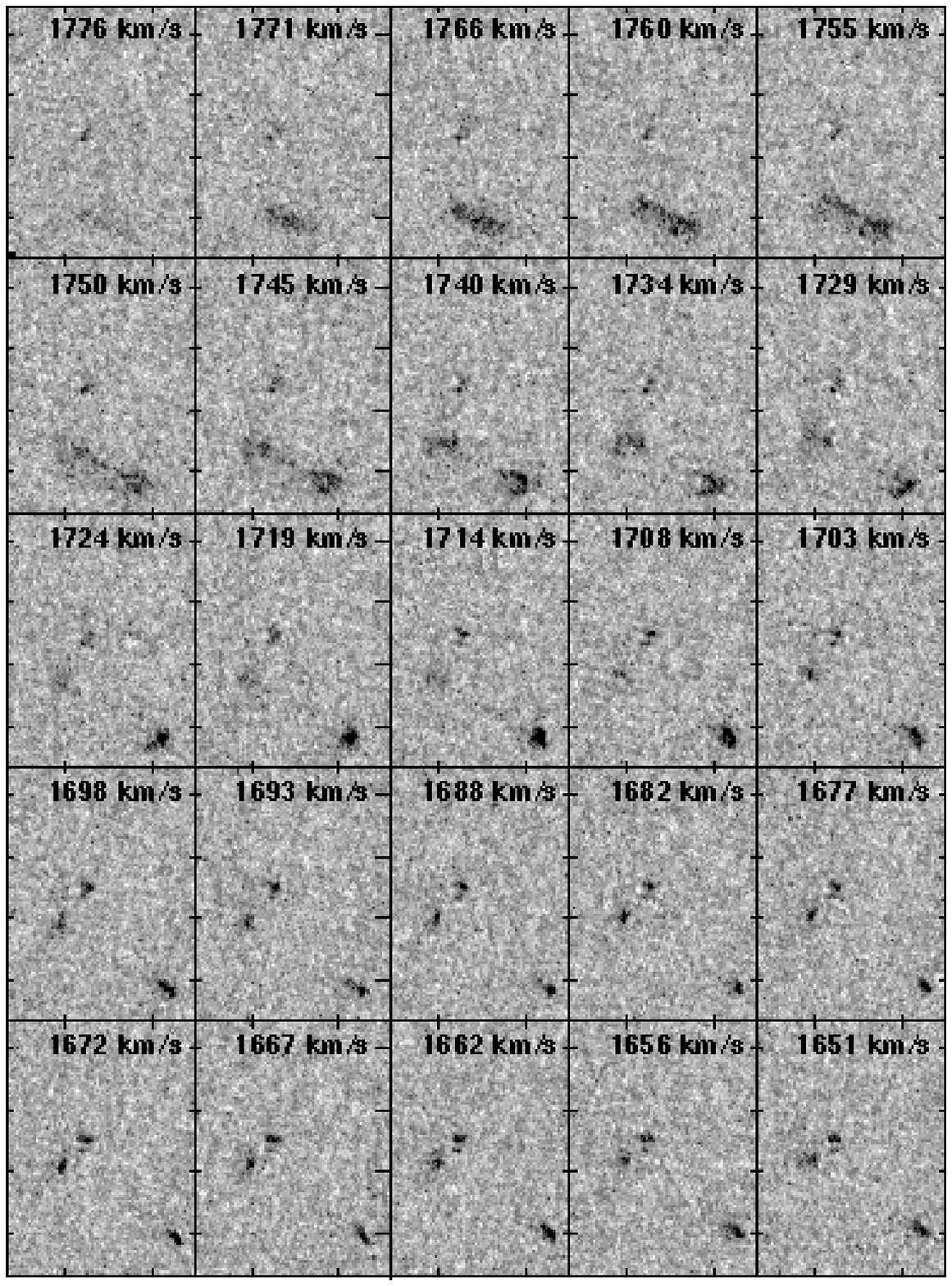}
\figcaption{
A greyscale representation of the intermediate resolution \hi\ channel
maps of the tidal regions of NGC 4038/9.  The greyscales are saturated
at 1.1\col{20}.  Each panel is labeled with its heliocentric
velocity.  Only channels showing {\hi} emission from the tails
are shown. This figure emphasizes the wealth of gaseous structure
present in the tidal regions.
\label{fig:greychan}}
\end{figure*}

The tail structure and kinematics are further illustrated in the
channel maps, shown in Figure~\ref{fig:ALLchan} (low resolution data
after Hanning smoothing by a factor of two in velocity to $\Delta
v$=10.6 \kms, contoured upon the optical) and Figure~\ref{fig:greychan}
(intermediate resolution data; $\Delta v$=5.2 \kms, shown as a
greyscale).  Only channels containing emission from the tidal tails
are shown in these figures, as the disk emission is spread over a much
broader range of velocities and are shown separately below (\S
\ref{sec:disk}).  Fig.~\ref{fig:ALLchan} gives a more complete mapping
of the gas within the tidal tails and shows its relationship to the
underlying starlight, while Fig.~\ref{fig:greychan} allows a more
detailed view of the structure of the gas within the tails.  

The greyscale representation of the channel maps in
Fig.~\ref{fig:greychan} clearly shows the ``bifurcated'' morphology of
the southern tail mentioned above.  The bifurcation is particularly
apparent in the channels between 1760 and 1734 \kms.  The two
filaments merge together in a sideways ``V'' shape in the vicinity of
the star forming regions identified by MDL92 (panel at 1734 \kms\ in
Fig.~\ref{fig:greychan}). The outer (southern) filament is associated
with the previously described high \MhLb\ material that lies off of
the optical tail (see also panels at 1729 -- 1772 \kms\ in
Fig.~\ref{fig:ALLchan}).  The inner (northern) filament projects onto
the optical tail and has a higher characteristic column density.

\begin{figure*}
\plotone{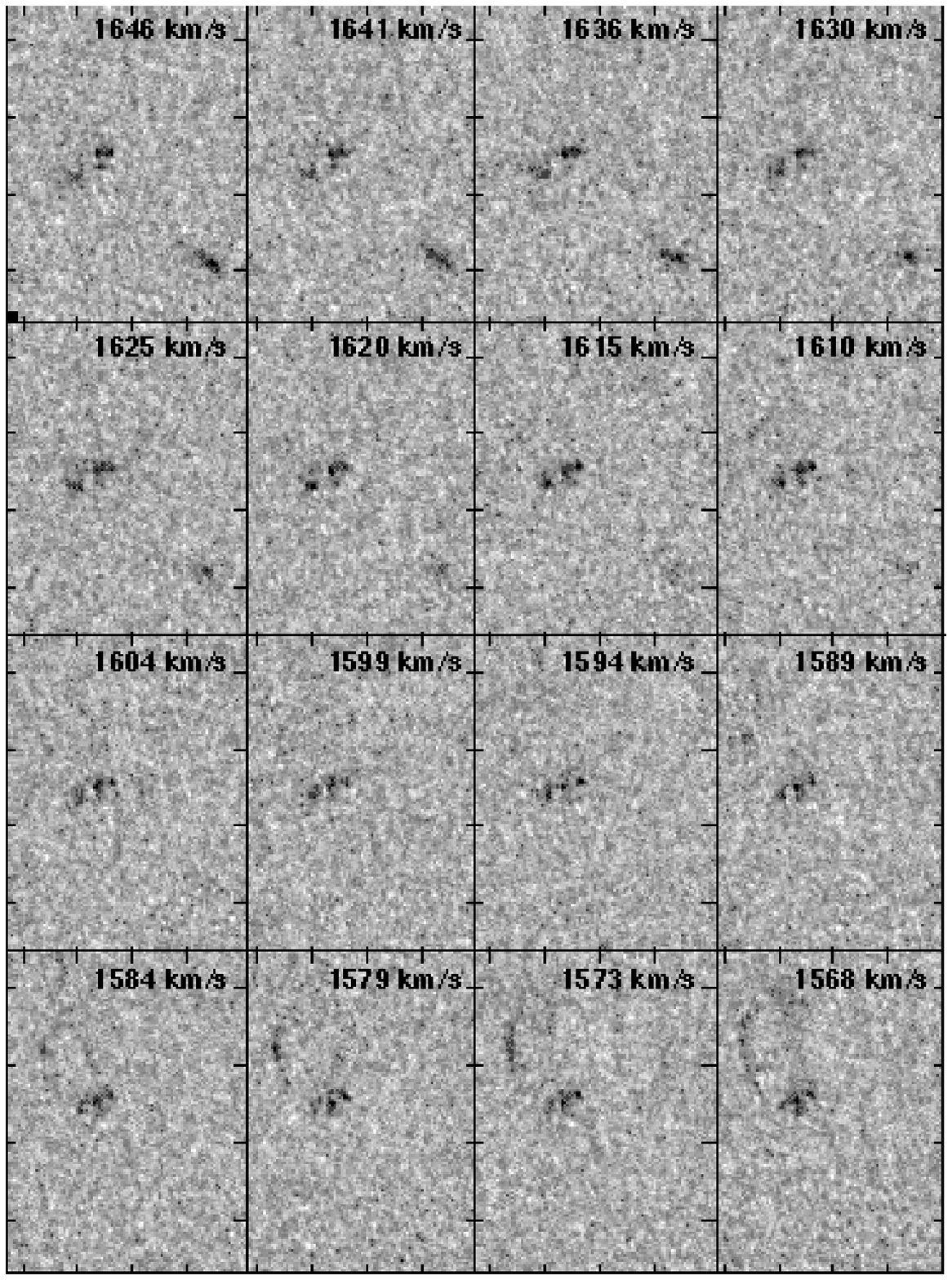}
\end{figure*}

\bigskip
\bigskip
\subsection{Northern Tail Morphology}
\label{sec:Ntail}

The low resolution moment map (Fig.~\ref{fig:HIdss}) shows that the
northern \hi\  tail extends beyond the end of the optical tail, to a
projected radius of 7\farcm3 (40 kpc) from the center of NGC
4039. This extension is confirmed by the ATCA observations of Gordon
\et\  (2001). We detect a total of 4.2\Mo{8} of \hi\  associated with 
the northern tail, which is much more than found in any of the
previous observations. The northern tail is redder
(Fig.~\ref{fig:ALLmos}b), less luminous, and has a lower relative \hi\
content than the southern tail (Table~\ref{tab:global}).

We also notice that the gas in the northern tail is relegated to the
outer half of the optical tail and in a purely gaseous extension
beyond this.  In particular, there is a distinct gap in the \hi\
distribution from a projected radius of 25 kpc back to where the tail
connects onto NGC 4039.  The difference in the \hi\  content of the
tails appears to be mirrored in the quantities of \hi\  within the
disks associated with either tail (with NGC 4038 being gas rich and
NGC 4039 being gas poor; see \S \ref{sec:disk} below).  These
differences and the lack of \hi\  at the base of the northern tail were
already noted by van der Hulst (1979a), who attributed them to a
difference in the \hi\  content of the parent galaxies.  We will return
to this point in \S
\ref{sec:ionize}

\subsection{Global Tail Kinematics}
\label{sec:TailKin}

Velocities along the northern tail show a regular gradient from 1610
\kms\  at the base of the \hi\  to 1540 \kms\  at the end of optical
tail, while the velocities along the southern tail range from 1590,
where the tail connects onto the northern disk, to a peak of 1770
\kms\  midway along the tail, declining to 1610 \kms\  toward the tip.
This kinematic continuity is convincingly illustrated in the
position-velocity plots in Fig.~\ref{fig:posvel}, which also
demonstrates that the tidal gas is quite dynamically cold. The smooth
run of velocities along the tails is already apparent from the
Westerbork data of van der Hulst (1979a), and is expected from 
tidal interaction models (e.g. Barnes 1988, Hibbard \& Mihos 1995).

The well-behaved kinematics enable us to infer the approximate tail
geometry. For example, along the southern tail the \hi\ column density
is lowest at the regions of the most extreme redshift, and increases
to either side of this. This indicates that this region of the tail
lies perpendicular to our line-of-sight, with its motion directed most
nearly toward us. At this location, the tail has a minimum projected
thickness and hence lowest \hi\ column density.  To either side of
this the tail curves toward our line-of-sight, resulting in a larger
pathlength through the gas-rich tidal material and hence a higher
projected column density. The sharper velocity gradients to either
side reflect the slewing of the velocity vectors as the tail bends
away from us.  Similarly, the sharp up-turn at the end of the southern
tail in both space and velocity suggests that this region curves away
from us. We will address the implications of this geometry for 
any embedded tidal dwarf galaxy in \S \ref{sec:TDG}.

Since tails are kinematically expanding structures (Toomre \& Toomre
1972, Barnes 1988), the relative velocities of the two tails indicate
that the southern tail is moving away from us and connects back to NGC
4038 from behind while the northern tail is swinging slightly toward
us and connects back to NGC 4039 from the front. This geometry is
supported by the lack of a dust absorption feature associated with the
southern tail as it crosses from the south back to NGC 4038. In fact,
the optical colors of these regions are quite blue
(Fig.~\ref{fig:ALLmos}b), whereas we would expect red colors from dust
absorption if this gas-rich material lies between us and the disk.  As
shown directly by van der Hulst (1979a), these kinematics are
consistent with the prograde spin geometry derived by the early
numerical models of this system by Toomre \& Toomre (1972; see also
Barnes 1988; Mihos, Bothun \& Richstone 1993).

Following the southern tail kinematics back to its progenitor disk
(NGC 4038, to the north), we see that the velocities become
blueshifted toward the base of the tail (Fig.~\ref{fig:posvel}).
Since the predominantly redshifted velocities of the tail constrain it
to be swinging away from us and connect back to the disk of NGC 4038
from behind, the blueshifted velocities at the base indicate material
that is streaming back toward the disk from the tidal tail (Hibbard
\et\  1994).  Such streaming is a natural consequence of such
encounters (e.g. Barnes 1988, Hernquist \& Spergel 1992, Hibbard \&
Mihos 1995), and arises because the tail material at smaller radii is
more tightly bound than material further out. As a result, this
material reaches its orbital apocenter more quickly and subsequently
falls back toward the inner regions.

We can use the maximum projected tail length of 65 kpc and the disk
rotation speed inferred from the Fabry-Perot observations of Amram
\et\  1992 (155 \kms, corrected for disk inclination), to infer an
interaction age (measured from orbital periapse, when the tails were
launched) of 420 Myr. This is in excellent agreement with the
best-match time of 450 Myr from the numerical model of NGC 4038/9 run
by Barnes (1988). It is also remarkably close to the age of the
population of 500 Myr old globular clusters found by Whitmore \et\
(1999; see also Fritze-v.Alvensleben 1998), supporting the hypothesis
that this GC population formed around the time when the tails were
first ejected (Whitmore \et\ 1999).

From Fig.~\ref{fig:ALLmos}d we note that the velocity dispersion
within the tidal tails is actually quite low: except for the base of
the southern tail, the \hi\ dispersion is less than 16
\kms, and the mean of 7 \kms\ is typical of values measured in
undisturbed disk galaxies. This is true of long-tailed mergers in
general (Hibbard \& van Gorkom 1996, Hibbard \& Yun 1999a). The
dispersion near the base of the southern tail has higher
characteristic values of 13--25 \kms. These regions likely lie along a
steeper gradient in the potential, and are also regions where the
line-of-sight velocities are changing rapidly (Fig.~\ref{fig:ALLmos}c). 
These effects will lead to larger real and apparent velocity gradients, 
increasing the single-beam dispersion.

\medskip
\subsection{The Tidal Dwarf Galaxy Candidates}
\label{sec:TDG}

\begin{figure*}[t!]
\plotone{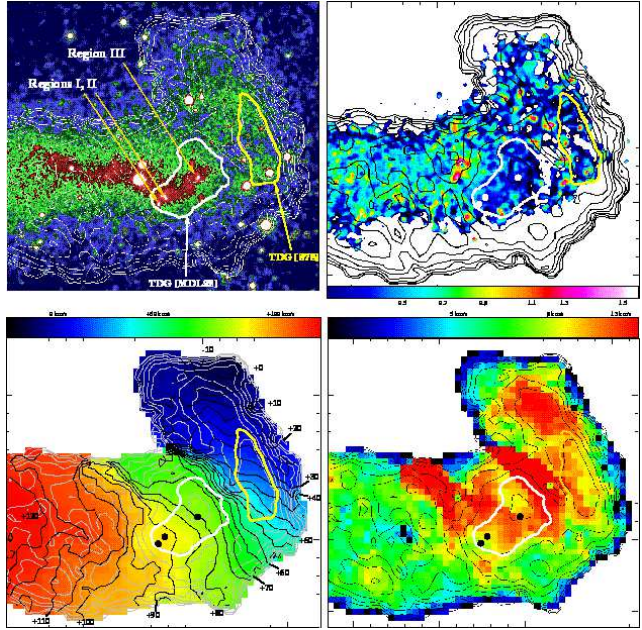}
\figcaption{
Four-panel mosaic of the region of the putative Tidal Dwarf
Galaxy. The upper left panel shows the optical image in false color,
with contours from the intermediate resolution \hi\ data drawn at
levels of 14.44 Jy \kms\ (5\col{19}) $\times 1.5^n$, $n=1-9$.  These
contours are repeated in all panels. The three star forming regions
identified by Mirabel \et\ (1992) are labeled in the upper left panel,
and are indicated by black or white dots in subsequent panels. Solid
heavy contours delineate the approximate half-light level around the
Tidal Dwarf candidates identified by Mirabel \et\ 1992 (heavy white
contour labeled TDG [MDL92]) and by Schweizer 1978 (heavy yellow
contour labeled TDG [S78]).  The upper right panel shows a color
representation of the $B-R$ color map, with the same color mapping as
in Fig.~\ref{fig:ALLmos}b.  The lower left panel shows the intensity
weighted \hi\ velocities with isovelocity contours drawn in black at 5
\kms\ intervals. The lower right panel maps the \hi\ velocity
dispersion, with the same color mapping as in Fig.~\ref{fig:ALLmos}d.
\label{fig:TDGmos}}
\end{figure*}

The improved resolution of the VLA observations allows us to directly
address the behavior of the atomic gas in the vicinity of the putative
tidal dwarf galaxy(s) identified near the end of the southern tail of
The Antennae. We show a close-up of this area in
Figure~\ref{fig:TDGmos}, which has the same arrangement as the panels
in Fig.~\ref{fig:ALLmos}.

Schweizer (1978) was the first to identify four \hii\ regions and blue
$UBV$ colors associated with the end of the southern tail. He also
identified a patch of low surface brightness material bending sharply
to the north after the end of the optical tail, which he suggested was
a separate dwarf stellar system. We indicate the approximate location
of this candidate TDG in Fig.~\ref{fig:TDGmos} using a thick yellow
contour drawn at approximately one half of the local peak \hi\ column
density after background subtraction and labeled TDG [S78]. Schweizer
noted that in the low-resolution \hi\ map of van der Hulst (1979), the
\hi\ appears to be more closely concentrated in the low surface
brightness extension of the tail, and hypothesized that the dwarf may
have been created during the interaction, as envisioned by Zwicky
(1956).

The end of the tail was subsequently studied by Mirabel et al. (1992),
who concentrated on what they called a ``detached condensation of gas
and stars at the tip of a tidal tail'', consisting of a twisted
stellar bar embedded in an envelope of diffuse optical emission and
containing three emission line complexes. We indicate the approximate
location of this candidate TDG in Fig.~\ref{fig:TDGmos} using a thick
white contour drawn at approximately one half of the local peak \hi\
column density after background subtraction and labeled TDG
[MDL92]. The three star forming regions identified by MDL92 are
labeled as Regions {\cap i--iii} in Fig.~\ref{fig:TDGmos}a, and are
represented by black or white dots in the remaining panels.

While all subsequent observers have referred to the region identified
by MDL92 when discussing the putative TDG, we will address the
dynamical nature of both of these regions. This is not a trivial task,
as the precise boundaries of the TDG candidates have never been
explicitly defined.

Our \hi\ observations (as well as those by Mahoney \et\ 1987 and
Gordon \et\ 2001) are a vast improvement on the observations of van
der Hulst (1979), and show that the tail is a continuous kinematic
structure. Specifically, there is no ``detached'' region which can
readily be identified as a distinct entity. Further, the velocity
field varies smoothly through the region containing the high gas
column densities and star forming regions without a significant twist
or kink.  The strongest velocity gradient corresponds with the region
where the tail turns abruptly to the north (compressed contours near
the center of the map in Fig.~\ref{fig:TDGmos}c, with a similar
gradient just to the east of this).  This smooth gradient suggests
that the velocity field in this region is dominated by projection
effects, with the tail bending away from us back into the plane of the
sky; as a result the velocity vector slews from pointing toward us to
pointing away from us. Therefore, we find no clear kinematic evidence
in the velocity field for any dynamically distinct entities. It is
possible that the strong tidal gradient may mask the kinematic
signature of any mass concentration.

Since there are no well defined boundaries to the dwarf candidates, we
calculate the \hi\ mass and blue luminosity in their vicinity by
summing the emission within successive circular apertures centered on
the local \hi\ column density peaks. The maximum aperture radius is
set by the half-width of the tail. A mean background is subtracted,
and the \hi\ flux is summed within each aperture of radius $R$. The
optical luminosity is measured from the smoothed star-subtracted
$B$-band within the same apertures.

\begin{figure*}[t!]
\plotone{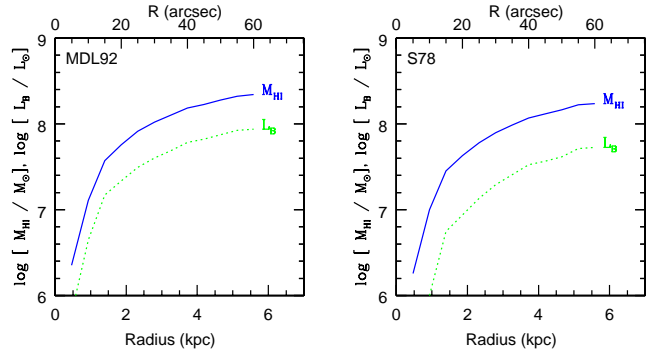}
\figcaption{Results of luminous mass calculation for the candidate 
Tidal Dwarfs identified by Schweizer 1978 (right panels, labeled TDG 
[S78]) and by Mirabel \et\ 1992 (left panels, labeled TDG [MDL92]). On 
both cases, the origin is taken to be the peak gas surface density in the
integrated intensity map (see Fig.~\ref{fig:TDGmos}). The two curves
show the enclosed \hi\ mass (solid curve) and optical luminosity (dashed 
curve) as a function of the circular aperture radius $R$. 
\label{fig:sumIRING}} 
\end{figure*}

The results of these calculations are shown in
Fig.~\ref{fig:sumIRING}. The left panel shows the result for TDG
[MDL92] and the right panel shows the result for TDG [S78]. In these
plots, the solid and dashed lines show how the \hi\ mass and blue
luminosity grow as a function of the aperture radius $R$.  In
Table~\ref{tab:TDG} we list the total \hi\ mass and optical luminosity
measured by these curves for each dwarf candidate, but since there are
no distinct boundaries for either region, these values should be
considered very rough guides. 

The strong velocity gradients mentioned above give rise to the regions
with the highest \hi\ velocity dispersions in Fig.~\ref{fig:TDGmos}d,
as gas with different space velocities fall within the same
beam. Aside from these regions, there are other local peaks in the gas
velocity dispersion which might provide possible evidence for a mass
concentration: between Regions {\cap i \& ii} and Region {\cap iii}
the \hi\ velocity dispersion increases to $\sim$12 \kms\ vs. an
average of 6--7 \kms\ along the tail.  Similar signatures are seen at
the location of tidal dwarf candidates within the optical tails of the
merger remnants NGC 7252 (Hibbard \et\ 1994) and NGC 3921 (Hibbard \&
van Gorkom 1993, 1996). However, it is noteworthy that the three \hii\
regions actually fall on the edges of \hi\ maxima. Similar signatures
are seen near giant \hii\ regions in dwarf galaxies (e.g. Stewart \et\
2000, Walter \& Brinks 1999), where the increased \hi\ linewidth is
due instead to kinetic agitation of the gas from energy deposited by
young stars, SNe and stellar winds (e.g. Stewart \et\ 2000, Yang \et\
1996, Tenorio-Tagle \& Bodenheimer 1988 and references therein) rather
than the gravitational effects of a mass concentration.

There is also an increased dispersion to the north of TDG [S78]
(dispersion increases to 15.6 \kms\ over the surrounding value of 11
\kms).  However, an examination of the line profiles shows that the
increased dispersion comes from low-level emission spread over many
channels. Adjacent regions have this broad low-level component, but 
also have a brighter narrower component which leads to a lower
intensity-weighted line-width. So we do not believe that this is a
signature of a mass concentration at this location. 

Since the moment maps are susceptible to line-of-sight integration
effects, we turn to the velocity cubes directly for further insight.
From the channel maps, we find that the gas near the TDG obtains the
highest gaseous phase-space densities in this system (i.e.~density
enhancements localized in space and velocity).  In fact, this region
has a higher \hi\ column density per unit \kms\ than any other region
in this system, including regions within the disk or overlap region
(see also Fig.~\ref{fig:posvel}a).  The peak of 21 mJy beam$^{-1}$ in a
single velocity channel of width 5.2 \kms\ corresponds to a single
channel column density of 3.8\col{20}, and the integrated column
density of 1.3\col{21} represents an enhancement of 6 over the mean
tail density. The high-resolution data shows an even higher
single-channel peak of 6.4\col{20}.

\begin{figure*}
\plotfiddle{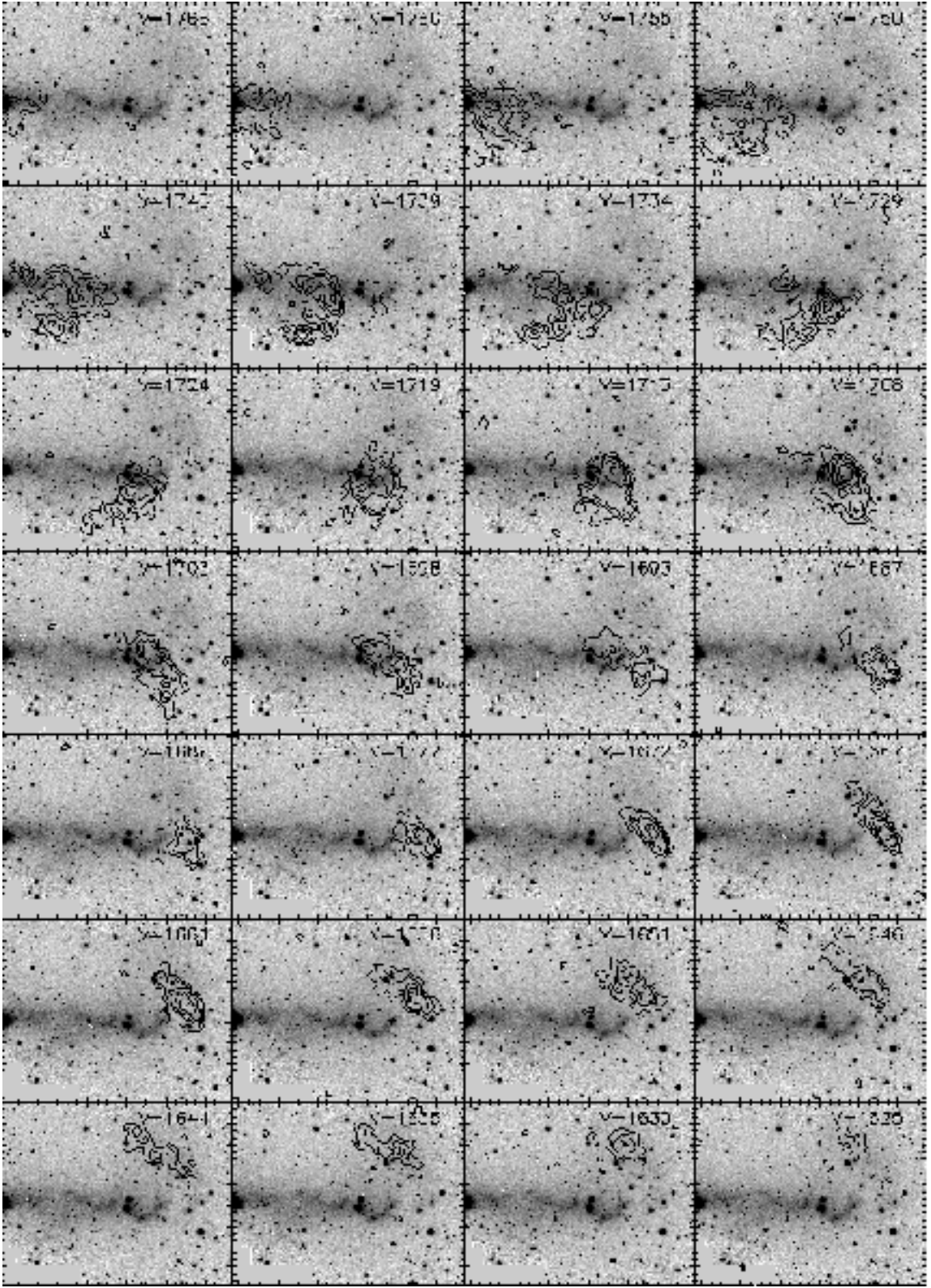}{19.5cm}{0}{100}{100}{-290}{-90}
\figcaption{
\hi\  channel maps of the region around the putative Tidal Dwarf Galaxy
within the southern tail of NGC 4038/9 contoured upon a greyscale
representation of the $B$-band image.  The greyscales range 
from \mB=28.5 \msqas\  (white) to 25 \msqas\  (black).  The
intermediate resolution \hi\  cube is used at its full velocity
resolution (5.2 \kms\  channel width).  Contours are drawn at levels of
2.8 mJy beam$^{-1} \times 1.5^n, n=0-4$, where 2.8 mJy beam$^{-1}$ 
corresponds to a single channel column density of 5\col{19}.
\label{fig:TDGchan}}
\end{figure*}

A closer examination shows that there are actually two dense knots in
this region.  These are seen in the channel maps from the intermediate
resolution datacube, plotted at full velocity resolution in
Fig.~\ref{fig:TDGchan}.  The densest concentration appears at $V$=1708
\kms\  and lies just north of star forming Region {\cap iii}. The second
concentration encompasses star forming Regions {\cap i \& ii} and
appears at $V$=1719--1729 \kms.  Such concentrations suggest regions
of gaseous dissipation.  While this is not unequivocal evidence of a
self-gravitating dwarf-sized object, since increased gaseous
dissipation is expected in regions of recent star formation, it does
indicate that there is something unique about this region of the tail.

The channel maps also show the bifurcated structure mentioned in \S
\ref{sec:Stail}. It is very interesting that the two parallel
filaments join just east of the location of the star forming regions
associated with the TDG (the ``V"-shaped feature seen in the panel
at 1734 \kms\ of Fig.~\ref{fig:TDGchan}).  Again, these observations
suggest that there is something special about the gas and tail
geometry at the location of the star forming regions.

The parallel filaments can be seen separately in the channels from
V=1755--1724 \kms.  The northern filament quite clearly lies along the
optical tail, while the southern filament is displaced by about 8 kpc
to the south.  The density peaks are of similar magnitude in either
filament, but only at the location of the putative TDG is there a
corresponding increase seen in the optical light underlying the gas
peaks.  The gas concentration associated with Region {\cap iii}
appears to lie within the northern filament, while it is difficult to
say whether the concentration associated with Regions {\cap i \& ii}
are associated with one filament or the other.  At velocities below
1693 \kms, it appears that only the southern filament continues on to
the gas rich ``hook'' at the end of the tail.

Fig.~\ref{fig:ALLmos}b shows that the blue tail colors discovered by
S78 are not restricted just to the high column density gas near either
TDG, but represents a trend of increasing blueness with distance along
the tail.  We find that while the outer disks and southern tail have
similar $B-R$ colors, the end of the southern tail is bluer in $V-R$
than $B-V$ while the reverse holds for the outer disks.  This suggests
that the tail material is younger and/or more metal poor than the
outer disk material (Schombert \et\ 1990, Weilbacher \et\ 2000),
although we postpone to a later paper a full color analysis (Evans et
al.\ 2001).

\medskip
\subsection{Disk Morphology and Kinematics}
\label{sec:disk}

\begin{figure*}[t!]
\plotone{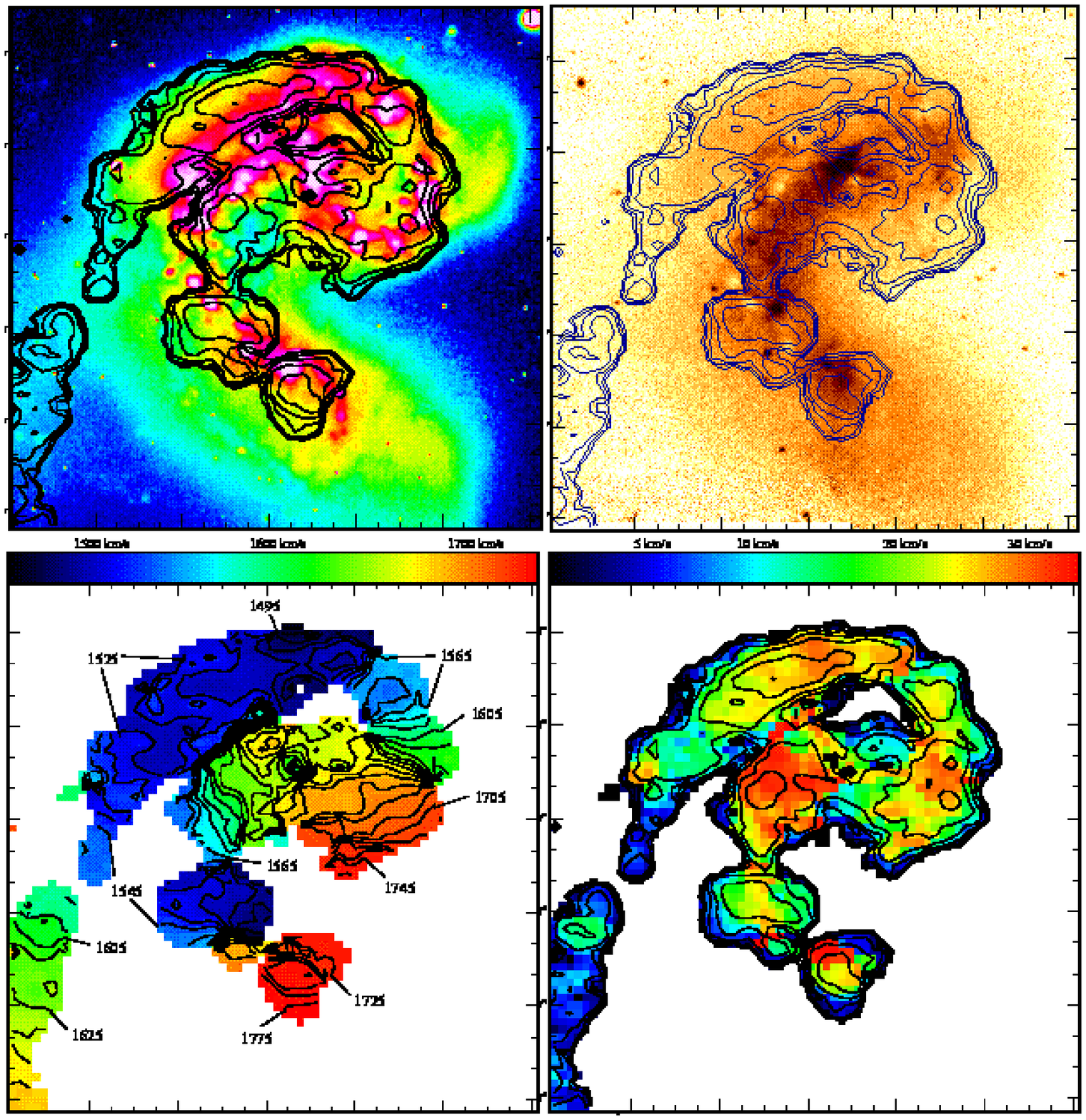}
\figcaption{   
Details of the \hi\  emission within the disks of NGC 4038/9.
The upper left panel shows contours from the high resolution 
\hi\  datacube ($11.4''\times 7.4''$) upon a false-color 
representation of the $B$-band image. Contours are drawn at levels of
(1, 2, 4, 8, ...) \x 7.74 mJy beam$^{-1}$ \kms, where the lowest contour
corresponds to a column density of 1\col{20}.  The upper right panel
shows the same \hi\ contours upon the $B-K$ color map, where lighter
shades indicate bluer colors and darker shades represent redder
colors. The middle two panels show the intensity weighted \hi\
velocities with isovelocity contours drawn at 5 \kms\
intervals (left) and the \hi\ velocity dispersion with the 
same \hi\ contours as in the first panel (right). 
The next two panels shows $K$-band image with the
\hi\ contours superimposed on the left, and the a
greyscale representation of the \hi\ image on the right, 
with the CO contours of Wilson \et\ (2001) superimposed 
and red crosses marking 
the location of the Super Star Clusters identified by
Whitmore \& Schweizer (their Table 4). 
\label{fig:DISKmos}}
\end{figure*}

Fig.~\ref{fig:DISKmos} shows an overlay of the high resolution \hi\
contours upon a false-color representation of the $B$-band image
(Fig.~\ref{fig:DISKmos}a), of the $B-K$ color map
(Fig.~\ref{fig:DISKmos}b), of the \hi\ velocity dispersion map
(Fig.~\ref{fig:DISKmos}d), and of the NIR $K$-band image
(Fig.~\ref{fig:DISKmos}e). In Fig.~\ref{fig:DISKmos}c we show a color
representation of the \hi\ velocity field with isovelocity contours
drawn at intervals of 5 \kms. Finally, in Fig.~\ref{fig:DISKmos}f we
show the integrated CO map of Wilson et al. (2000) contoured in green
upon a greyscale representation of the \hi\ column density. The latter
image gives a better idea of the local minima and maxima in the \hi\
distribution, as these are not always self-evident in the contour
map. In this panel red crosses indicate the location of the ``Super
Star Clusters" (SSCs) identified in the HST image of Whitmore et
al.~(1995, from their Table 1).

\begin{figure*}[t!]
\plotone{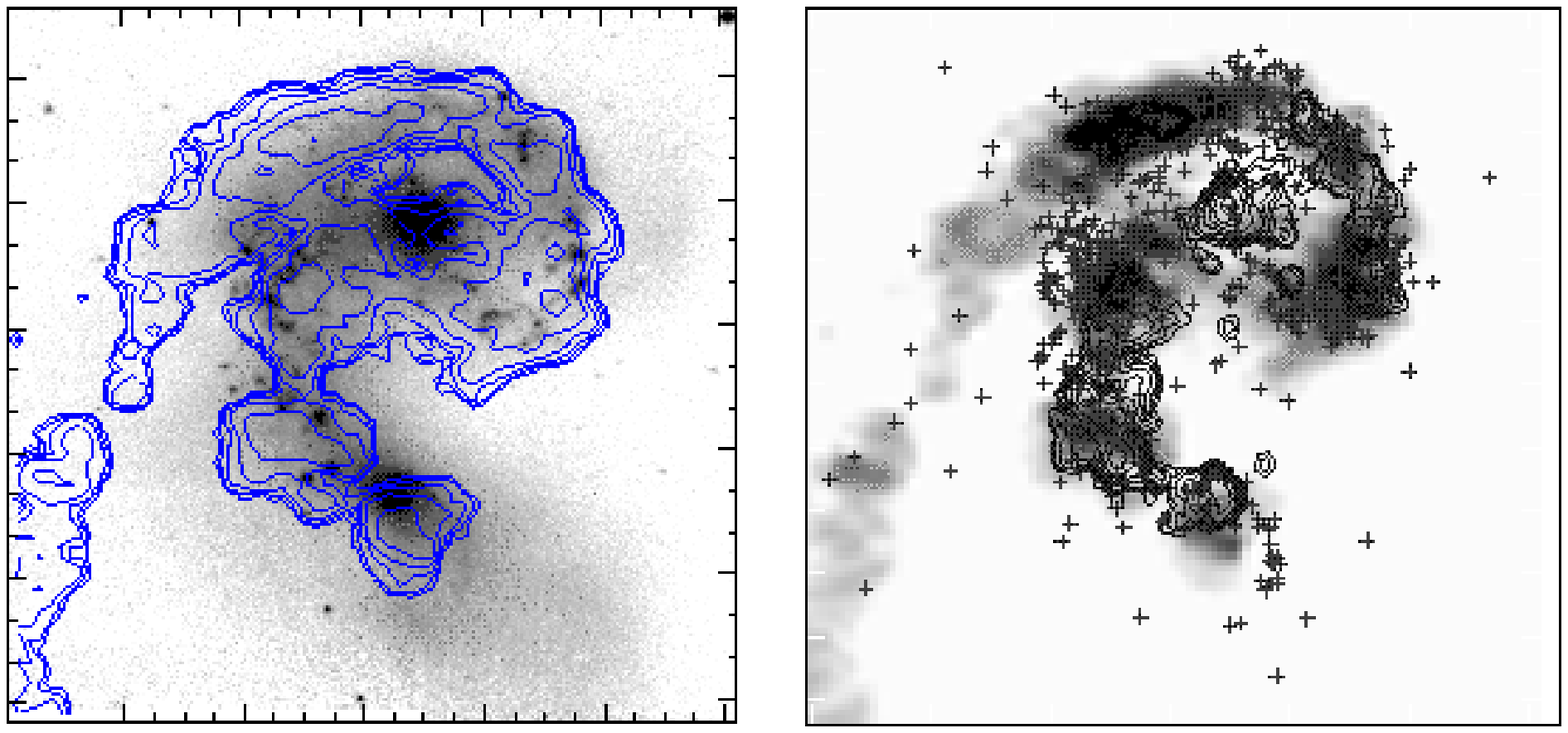}
\end{figure*}

In Fig.~\ref{fig:DISKmos}b dark regions represent the reddest $B-K$
colors, generally indicating regions of high dust extinction, while
light regions represent the bluest colors, generally indicating
unobscured massive young stars.  Comparison of Figs.~\ref{fig:DISKmos}b 
\& f shows that the CO is closely associated with the red features in 
the $B-K$ map, so that the CO traces the dust distribution. 

Figs.~\ref{fig:DISKmos}b \& f show that on a global scale, the \hi\
distribution traces both the regions containing the most star clusters
and the regions with the highest dust content, e.g.~the star forming
ring in NGC 4038 and the star forming half of the disk of NGC 4039 (to
the northeast of the nucleus), and the dusty region between the two
galaxies and to the southwest of NGC 4039.  On finer scales (i.e.~of
order a few kpc), the \hi\ is not particularly associated with the
bluest star clusters or reddest dust concentrations.  Most of the SSCs
in Fig.~\ref{fig:DISKmos}f actually appear to lie along steep
gradients in the \hi\ and CO distributions.  The molecular gas peaks
are often displaced from the $N_{HI}$ peaks and many of the SSCs
actually lie in the transition zone between the two gas phases. Where
there are no displaced CO peaks, the SSCs still lie predominantly
along steep gradients in the \hi\ distribution.

The NIR image in Figs.~\ref{fig:DISKmos}b does not show additional 
light peaks in the regions of highest \hi\ column density (recall
that dust extinction is 10 times weaker in the NIR compared to the
optical), so we conclude that the occurrence of SSCs along gradients 
in the \hi\ distribution is not due to dust associated
with the \hi\  obscuring clusters at the regions of highest
\hi\  column density.  Rather, we suspect that it is due to young stars
ionizing or sweeping the \hi\  on the one hand, and the conversion of
\hi\  to molecular form on the other.
We refer the reader to Zhang, Fall \& Whitmore (2001) for a much more
thorough discussion of the relationship between SSC's of various ages
and the \hi, Radio continuum, CO, X-ray, FIR, MIR, $H\alpha$ and
optical morphology and the \hi\ and CO kinematics.

The lower \hi\ contours in Fig.~\ref{fig:DISKmos} are very close
together, with the \hi\ column density falling off sharply below
4\col{20}.  The western part of the disk of NGC 4038 is largely devoid
of \hi, as is most of the extended disk of NGC 4039. Examination of
the lower resolution \hi\ maps (Figs.~\ref{fig:HIdss},
\ref{fig:ALLmos}a) shows that this is true even at lower \hi\ column
densities. Probably not coincidently, these regions lack current star
forming regions.

The intensity weighted \hi\ velocities of NGC 4038/9
(Fig.~\ref{fig:DISKmos}c) shows a very disturbed velocity field,
similar to those mapped in \hi\ (Gordon \et\ 2001) H$\alpha$ (Burbidge
\& Burbidge 1966, Rubin \et\ 1970, Amram \et\ 1992) and in CO
(Stanford \et\ 1990, Wilson \et\ 2000, Zhu 2001).  In particular, the
disk of NGC 4038 has a north-south gradient spanning $\sim$ 250 \kms,
while NGC 4039 has a northeast-southwest gradient spanning a similar
range (see also Fig.~\ref{fig:posvel}).  The general sense of motion
--- north blueshifted, south redshifted --- agrees with the sense of
rotation inferred from the tidal tail. However, because of the
extremely disturbed nature of both disks, the kinematics of these
regions are not simply rotational.  It is likely that there are strong
streaming motions along the star forming loop and along the overlap
region connecting the two disks.

\begin{figure*}
\plotfiddle{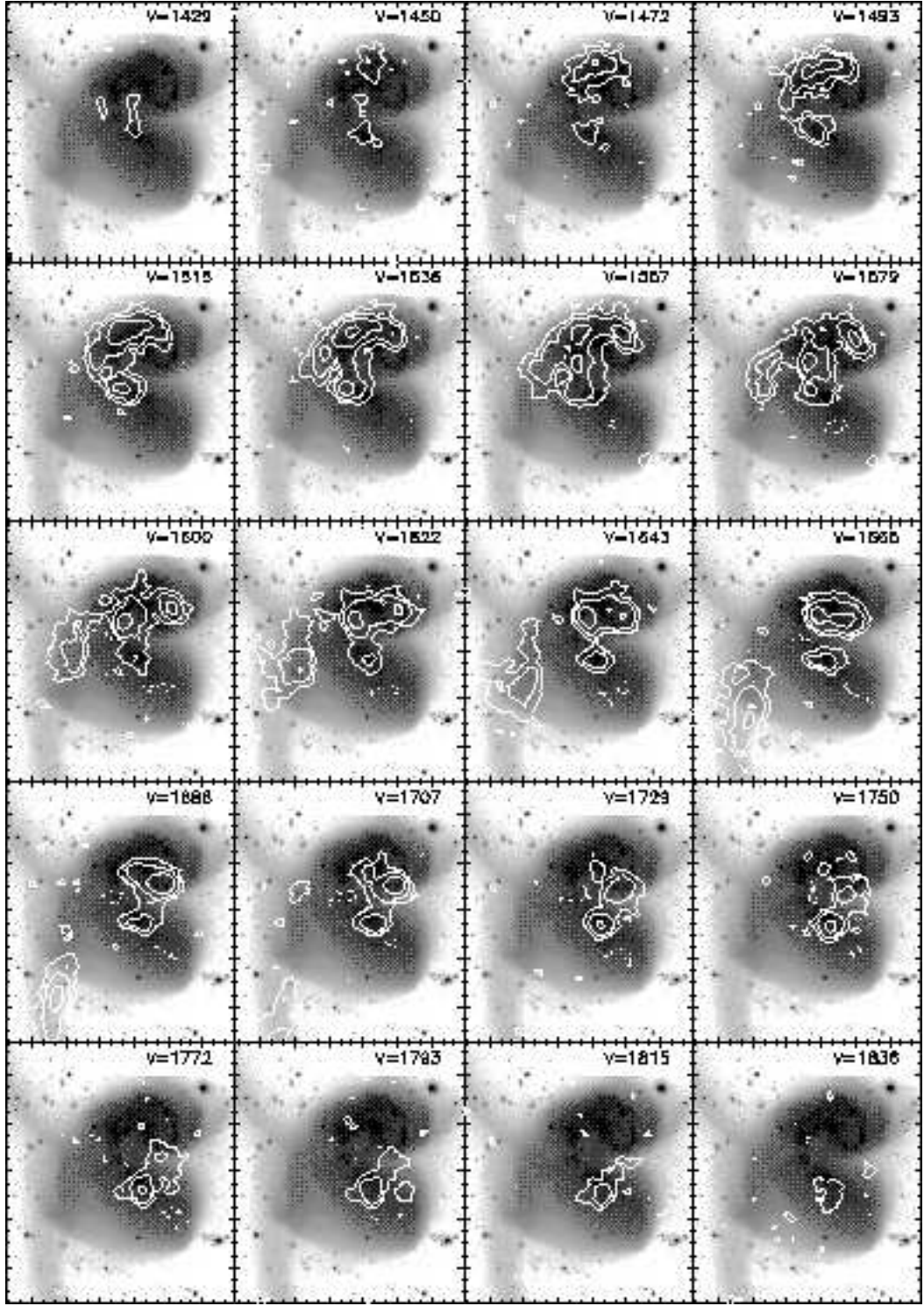}{19cm}{0}{100}{100}{-300}{-100}
\figcaption{
\hi\  channel maps of the disk regions of NGC 4038/9
contoured upon the CTIO $B$-band image.  The intermediate resolution
\hi\ cube is used after smoothing by a factor of four in velocity (to
a 21 \kms\ channel width).  Each panel is labeled with its
heliocentric velocity.  Contours are drawn at levels of (-3, 3, 6, 12,
24, ...) \x 0.4 mJy beam$^{-1}$, where 0.4 mJy beam$^{-1}$ is the
rebinned channel noise level, corresponding to a column density of
3\col{19}. These maps show the entire velocity spread of the present
observations, and the reader should refer to Gordon \et\ (2001) for
\hi\ emission from the disk regions that falls outside of our
bandpass.
\label{fig:DISKchan}}
\end{figure*}

In Figure~\ref{fig:DISKchan} we present the disk channel maps, made
from the intermediate resolution data after Hanning smoothing by a
factor of four in velocity ($\Delta v$=20.4 \kms). Instead of a normal
rotational pattern, the redshifted gas concentration to the SW of NGC
4039 shows very little spatial gradient over the velocity range
V=1836--1708 \kms, while the blueshifted gas to the northeast of NGC
4039 has very little spatial gradient over the velocity range
V=1579--1472 \kms. The gas not associated with either of these two
regions appears to flow from the NE at V=1558 \kms\ to the SW at
V=1665 \kms. This is likely material which is flowing along the
overlap region between the two galaxies.  The \hi\ in the northeast
part of NGC 4039 may be material which is being accreted from NGC 4038
along the bridge connecting the two systems.  We also notice that the
disk gas connects smoothly to the gas associated with the base of the
southern tail (channels at V=1472--1686 \kms). From the tail
kinematics (\S \ref{sec:TailKin}) we concluded that the blueshifted
velocities associated with the base of the tail require that this gas
streams back onto the outer disk of NGC 4038 from the tail. In
Fig.~\ref{fig:DISKmos}d we find a region of high velocity dispersion
along the northern loop of gas (dispersion of 28 \kms\ vs 22 \kms\ at
adjacent regions). We suggest that this increase is associated with
gas falling back from the tail mixing along the line-of-sight with gas
associated with the disk.

\subsection{The Nearby Dwarf Galaxy ESO 572--G045} 

As part of this observation we also detect \hi\  associated with the Scd
dwarf galaxy ESO 572--G045 just to the southwest of the tip of the
southern tail (5.7$'$ or 32 kpc from the TDG).  This system lies at a
projected distance of 90 kpc from The Antennae.  This galaxy is very
flattened, with an axis ratio near 10, associating it with the class
of ``superthin'' galaxies (Goad \& Roberts 1981), and qualifying it
for membership in the ``Flat Galaxy Catalogue'' of Karachentsev et
al.~(1993).  It is curious that a similar superthin galaxy was also
serendipitously discovered by Duc \et\ (2000) in their HI mapping of 
the interacting system NGC 2992/3, another interacting pair with a
candidate Tidal Dwarf within a gas-rich tidal tail.

\begin{figure*}
\plotone{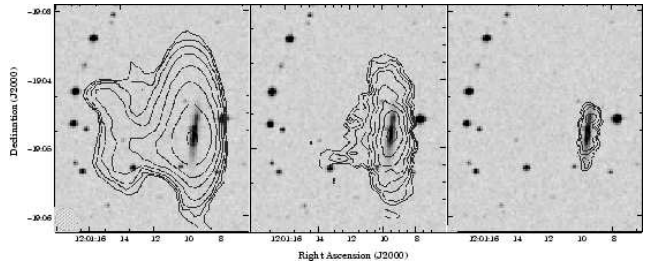}
\figcaption{Contours of the integrated \hi\  emission of the
small superthin galaxy ESO 572--G045 overlaid on the DSS image. These
panels show the \hi\  data from the low-- (left), intermediate--
(middle) and high-- (right) resolution datacubes, with the beam size
of each dataset drawn in the lower left corner. 
Contours are drawn at levels of [144.9 Jy \kms\ \x $2^{n/2}$, 
57.75 Jy \kms\ \x $2^{n/2}$, 18.1 Jy \kms\ \x $2^n$] ($n=0,1,2...$), for
the low, intermediate and high-resolution data, respectively.
The lowest contour corresponds to a column density of [1\col{20}, 
1\col{20}, 4\col{20}], respectively. 
Notice the gaseous extensions extending to the 
east in the intermediate and low resolution maps.
\label{fig:ESOmom0}} 
\end{figure*}

Contours of the \hi\ column density from the various resolution data
cubes are shown on the optical image from the DSS in
Figure~\ref{fig:ESOmom0}.  We detect a total 3.7\Mo{8} of \hi\ spread
over 100 \kms\ and extending to a radius of 6 kpc (i.e., twice as far
as the optical light). The galaxy is nearly edge-on and shows a linear
velocity gradient.  Assuming an edge-on orientation and circular
velocities (with $V_{circ}$ = 50 \kms), this suggests a dynamical mass
($M_{dyn}$) of 3.5\Mo{9} out to 6 kpc.  Using the optical properties
given in NED (major and minor axis of $1.01'\times0.11'$ and
$m_B=$16.70 mag), we derive dimensions of 5.9 kpc \x\ 0.6 kpc and a
luminosity of $L_B=1.9\times 10^8 L_\odot$. This system is quite gas
rich ($M_{HI}/M_{dyn}=0.11, M_{HI}/L_B=1.9$), with global properties
similar to those of other superthin galaxies (Matthews \et\ 1999).

The low resolution \hi\ map in Fig.~\ref{fig:ESOmom0}a clearly shows
some low column density gas extending to the east.  Since this
extension is one-sided in nature, it is more indicative of ram
pressure effects than tidal effects. The very thin and flat disk of
ESO 572--G045 also argues against significant tidal forces acting on
it.  This suggests that there may be an extended warm or hot gaseous
halo around NGC 4038/9 that is affecting the diffuse cold ISM of ESO
572--G045 (e.g. Moore \& Davies 1994).

\section{Discussion}
\label{sec:disc}

The wealth of information provided by these observations is a
testament to the utility of \hi\  spectral line mapping of disturbed
systems. While in some sense the present data are just the beginning
point of many more involved investigations, particularly for dynamical
modeling of this system, they also lead to many interesting results on
their own, which we discuss in this section.

\subsection{The lack of \hi\  at the base of the Northern Tail}
\label{sec:ionize}

The difference between the northern and southern tail in both optical
structure and \hi\  content was already noted by van der Hulst (1979a), 
who attributed it to a difference in the \hi\  content of the parent
galaxies. The lack of \hi\  within the main disk NGC 4039 (i.e. the
progenitor of the northern tail) was seen as support for this
interpretation. However, we now feel that the observations are
at odds with this explanation. For while it is true that the
outermost regions of tidal tails should originate from the
outermost regions of the progenitor disks, it is not true that the
outermost regions of the progenitor disks end up {\it only} at the
outermost regions of tidal tails. In fact, this gas rich material
should extend all the way back to the host system. This fact is
apparent already in the simple models and illustrations of Toomre \&
Toomre 1972 (see their Fig.~2). 
This reflects the fact that tails are not linear structures, but
actually two dimensional ``ribbons" twisting through space, with the
outer disk material forming a ``sheath" around the inner disk material
(Mihos 2001). As a result, the gas rich outer regions should extend
along the entire length of the tidal feature back toward the
progenitor disk.  It therefore seems that a low \hi\  gas content alone
cannot explain the gas-rich outer tail and gas-poor inner tail.

In a recent paper, Mihos (2001) describes a kinematically decoupling
between the gaseous and stellar components of a tail that occur during
tail formation. However, this decoupling leads to a displacement
between the gaseous and stellar tidal features, and does not totally
remove large quantities of gas from along radial segments of the tail.
It is unlikely, therefore, that this mechanism explains the present
observations.

Hibbard, Vacca \& Yun (2000) investigate several possible mechanisms
for creating differences between \hi\ and optical tidal morphologies.
Two plausible mechanisms are ram-pressure stripping of the tidal gas
from an expanding superwind and ionization of the gas by UV photons
escaping from the starburst.  While there is some evidence for a
nuclear outflow in NGC 4038/9 (Read, Ponman, \& Wolstencroft 1995,
Sansom \et\ 1996, Fabbiano, Schweizer \& Mackie 1997, Lipari \et\
2001, Fabbiano, Zezas \& Murray 2001), the soft X-ray morphology (Read
\et\ 1995, Fabbiano \et\ 2001) does not suggest that this wind is
directed toward the northern tail, so we concentrate instead on
possible ionization effects.

The situation explored by Hibbard \et\  (2000) is that the starburst
has cleared enough dust and gas from the inner regions to provide
direct unobscured sightlines to the tails. By equating the surface
ionization rate with the recombination rate, they find that the gas
should be ionized out to a distance given by:

\begin{eqnarray}
R_{ionized} & \leq & 25 \,{\rm kpc} \,
\left({f_{esc}\over 0.05}\right)^{1/2}
\left({L_{IR}\over 6\times10^{10} L_\odot}\right)^{1/2}\nonumber\\
& \times & \left({2\times10^{20} {\rm cm}^{-2} \over N_{HI}}\right) 
\left({dL \over 5 \,{\rm kpc}}\right)^{1/2}
\label{eq:Rion}
\end{eqnarray}

Where $N_{HI}$ and $dL$ are the column density and thickness of the \hi\
layer being ionized, $f_{esc}$ is the fraction of ionized photons
emerging from the starburst, and 
we have made the standard
assumption that most of the starburst luminosity is emitted in the far
infrared (see Hibbard \et\  2000 for details).

The fiducial values used in eqn.~\ref{eq:Rion} are
appropriate for NGC~4038/9, and we thus find that the starburst may
well be capable of ionizing gas out to the \hi\ base of the northern
tail (25 kpc projected radius), providing the tail has a clear
sightline to the ionizing stars and that $f_{esc}$ is not much smaller
than 0.05. Evidence that $f_{esc}$ is at least as large as this is
provided by numerical work by Dove, Shull \& Ferrara (2000), who
derive $f_{esc}\sim 0.1$ for normal galaxies, and suggest that
starbursts should have significantly higher values. Hurwitz, Jelinsky,
\& Dixon (1997) find upper limits to the escape fraction of $UV$
photons of 3--57\% for four IR luminous starbursts, while
Bland-Hawthorn \& Maloney (1999) calculate $f_{esc}=0.06$ for the Milky Way
(see discussion in Bland-Hawthorn \& Putman 2001).

So the question becomes, is it likely that the northern tail of NGC
4039 has a relatively clear view of the star-forming disk, while the
southern tail does not? We believe the answer to this question is yes.
The majority of the unobscured massive star forming regions in NGC
4038/9 are associated with the disk of NGC 4038 and the material along
the bridge (Whitmore \& Schweizer 1995, Whitmore \et\ 1999).  Since
tidal tails are spun off close to the spin plane of the progenitor
(Toomre \& Toomre 1972), the southern tail will not have a clear view
to the star forming regions located within the disk NGC 4038, and the
body of NGC 4039 would block ionization originating from the bridge
star forming regions.  The body of NGC 4039, on the other hand,
appears to have a perfect view of the star forming disk of NGC 4038
and the northern tail should have a view to the backside of this disk.
This prediction can be tested by numerical models to constrain the
space geometry of the tail.

We suggest that NGC 4039 had a normal \hi\  distribution, but that much
of this gas has been photoionized by the starburst.  We believe a
similar effect is responsible for the lack of \hi\  at smaller radii in
the otherwise gas-rich tidal tails of the NGC 7252, Arp 105, and Arp
299 systems (Hibbard, Vacca \& Yun 2000).
If this is indeed the case, then the ionized gas should be visible via
its recombination radiation. Using the equation given in Hibbard \et\
(2000), the expected emission measure of this radiation should be $EM
= 0.8 \, {\rm cm^{-6}\, pc}$ for a gas column density of 2\col{20} and
thickness of 5 kpc.  This is within the capabilities of modern
CCD detectors (e.g.\  Donahue \et\  1995; Hoopes, Walterbos \& Rand
1999).

\subsection{The Bifurcation of the Southern Tail}
\label{sec:bifurcation}

As mentioned above, tails are two dimensional ``ribbons" twisting
through space.  It is possible that a lateral twist may cause the
outer edge of the ribbon to lie in a different plane from the inner
regions, i.e.~for the most gas-rich (highest \MhLb) regions to appear
in a different plane from the less gas-rich but optically brighter
regions (Mihos 2001).  In Hibbard \& Yun (1999a) we suggest that this
this effect may be exacerbated by a pre-existing warp in the
progenitor disk, providing even more marked offsets.  This provides a
simple explanation for the gas-rich but optically faint outer edge of
the southern tail.  Similar bifurcated tails are seen in several
systems (e.g.~M81, van der Hulst 1979b, Yun \et\  1994; NGC 3921,
Hibbard \& van Gorkom 1996; NGC 2535/6, Kaufman \et\  1997; Arp 299 
Hibbard \& Yun 1999a).

What is intriguing is that the two filaments appear to join back
together just at the location of the star forming regions associated
with the candidate tidal dwarf TDG [MDL92] (the
``V''-shape seen in Figs.~\ref{fig:greychan} \& \ref{fig:TDGchan} and
discussed in \S \ref{sec:Stail} \& \S \ref{sec:TDG}). We do not know 
whether this is a coincidence or an important clue. Further insight 
into this question will have to await a more detailed numerical study 
of this system.

\subsection{The Formation of the Tidal Dwarf Galaxy}
\label{sec:TDGform}

\begin{figure*}[t!]
\plotone{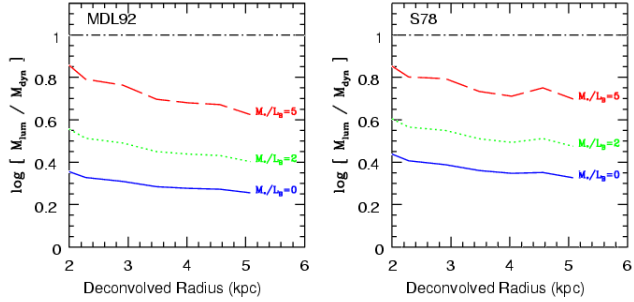}
\figcaption{Results of dynamical analysis for the candidate Tidal
Dwarfs identified by Schweizer 1978 (right panels, labeled TDG [S78])
and by Mirabel \et\ 1992 (left panels, labeled TDG [MDL92]). On both
cases, the origin is taken to be the peak gas surface density in the
integrated intensity map (see Fig.~\ref{fig:TDGmos}). We plot
$M_{lum}/M_{dyn}$ as a function of the deconvolved aperture 
radius $r$. Three curves are drawn, each one representing a different 
stellar mass-to-light ratio: the thick solid curve shows the result 
for $M_* / L_B = 0$ \ML, the dotted curve shows the results for 
$M_* / L_B = 2$ \ML, and the dashed line shows the results for 
$M_* / L_B = 5$ \ML.
\label{fig:MlumMdyn}} 
\end{figure*}

We wish to take advantage of the higher spatial and velocity
resolution of our \hi\ observations to examine the dynamical nature of
material in the vicinity of the TDG candidates.  This is a difficult
task, since there are no distinct objects which can be defined
unambiguously.  Further, if there are objects within the tail, they
are clearly not in isolation, and the kinematics may not be due to the
local mass concentration. All of these factors make a clear dynamical
analysis fraught with uncertainties. Still, the present data are the
best available, in terms of both spatial and kinematic
resolution, and this is the nearest system with a candidate tidal
dwarf, so we will proceed with all due caution.

The first hypothesis we will test is whether there are concentrations
of gas and light centered on the gas peaks with enough mass in
luminous matter alone to account for the measured \hi\ line-width. To
this end, we compare the virial mass inferred from the \hi\ linewidth
($M_{vir}$) to the total luminous mass inferred from the \hi\ flux and
the optical luminosity ($M_{lum}$). A value of $M_{lum}/M_{vir} = 0.5$
is sufficient for a region to be bound, while a value of
$M_{lum}/M_{vir} = 1$ is required for the region to be virialized.

To calculate the virial mass, we must make some assumptions about the
3-dimensional structure of any mass concentration. For the sake of
simplicity, we assume that there is an isotropic spherical mass
concentration of constant density. For this situation, the virial mass
is given by $M_{vir} = 3 \sigma^2 \times a r_h / G$, where $G$ is the
gravitational constant, $\sigma$ is the one-dimensional velocity
dispersion, $r_h$ is the projected half-light radius, and $a$ is a
geometric factor (Binney \& Tremaine 1987) $\simeq 2.74$ for the
adopted geometry. We take the velocity dispersion at the location of
the peak column density, and use the deconvolved half-light radius
found from fitting a Gaussian plus constant background  to the \hi\
mass distribution.

The luminous mass is calculated from the observed mass and optical
luminosity following the prescription given in Paper II.  In
particular, the possible contribution to the luminous mass by stars is
estimated by adopting a range of stellar mass-to-light ratios
($M_*/L_B$) representing different star formation histories
(SFHs). The gas mass is derived from the \hi\ mass by multiplying by a
factor of 1.36 to take into account a primordial abundance of Helium.
No correction is made for the presence of molecular gas since recent
CO observations at the location of the TDG by Gao \et\ (2001) and
Braine \et\ (2001) find only trace amounts of molecular gas in the
vicinity of the tidal dwarf (4\Mo{6}).

We evaluate three different stellar mass-to-light ratios, which span a
range of star formation histories. We evaluate $M_* / L_B =$ 0 \ML\
(i.e., $M_{lum}=M_{gas}$, also appropriate for a young instantaneous
burst); $M_* / L_B =$ 2 \ML, corresponding to a constant SFH at an age
of 10 Gyr, and $M_* / L_B =$ 5 \ML, representing an exponentially
decreasing SFH with a time constant of 4 Gyr at an age of 10 Gyr (see
Paper II for more details).  The results of the luminous to virial
mass calculation are given in Table~\ref{tab:TDG}.

The ratios of the luminous to virial mass given in Table~\ref{tab:TDG}
are less than one, suggesting that neither mass concentration is 
virialized. However, they do not rule out the presence of a bound
object.  They do imply that there is not a sufficient gas density for
the regions to be bound by gas alone (i.e.,
$M_{lum}/M_{vir}\,<\,0.5$ for $M_* / L_B =$ 0 \ML). This contradicts
the conclusion reached for this same region by Braine \et\ (2001). The
reason for this discrepancy is that Braine \et\ use the much narrow CO
linewidth to calculate a dynamical mass, yet include both \hi\ and
stellar light in their luminous mass calculation (as mentioned above,
the mass in molecular mass inferred from the CO line flux is negligible
compared to the \hi). The fact that the \hi\ linewidth is much broader
than the CO linewidth argues that a much larger mass concentration is
required to bind it, and the Braine analysis is inappropriate. This
point is discussed in more detail in Paper II.

A further conclusion from the above analysis is that if the dwarf
candidates {\it are} bound, they require a significant contribution to
the total mass from evolved stars, or they are bound by dark
matter. In this case, the calculated values of $M_{lum}/M_{vir}$ are
valid {\it only} if the \hi\ line-width also characterizes the motion
of the stars (and/or dark matter). This is probably not the case, since
in normal disks the older stellar populations (i.e., those with higher
$M_* / L_B$ ratios) have a significantly higher dispersion than the
gas, by factors of 2--3 (van der Kruit 1988), which would lead to a
virial mass estimate a factor of 4--10 higher and $M_{lum}/M_{vir}$
ratios correspondingly smaller. Young stellar populations have a
velocity dispersions similar to the cold gas, but also have smaller
$M_* / L_B$ ratios. A measurement of the stellar kinematics and a
better estimate of the stellar mass-to-light ratio is needed to be
more conclusive.  Finally, we note that the expected circular velocity
given by $V_c^2 = G M_{lum} / R$ is 20--30
\kms. Fig.~\ref{fig:TDGmos}c shows no sign of rotation at this
level. We conclude that either TDG candidate may be, but need not be,
self-gravitating.

This virial analysis includes many caveats. Primarily, the above
analysis implies that we know the ratio $M_{lum}/M_{vir}$ to within a
factor of two, and in truth we do not know either quantity to this
precision.  Further, unless the bound regions are much smaller than
the width of the tail, the geometry is unlikely to be spherical, and
the velocities are unlikely to be isotropic. Finally, any entrained
objects are clearly not in isolation, complicating the virial
analysis.  We will be addressing these uncertainties using simulated
observations of bound objects within N-body tails (Kohring, Hibbard \&
Barnes, in preparation).

The above analysis depends on a specific geometry and the details of
the Gaussian fit to the gas distribution. A more common but even less
clear analysis uses a dynamical mass indicator $M_{dyn} \sim \Delta V^2
\times R / G$, where $\Delta V$ is the FWHM linewidth, $R$ is a
radius, and $G$ is the gravitational constant (e.g. Braine \et\ 2001).
This quantity has the dimensions of mass, but it is not obvious what
mass it defines, especially when the radius used is not physically
motivated. However, we can use the quantity to see how taking larger
or smaller region in the vicinity of either dwarf candidate might
affect our conclusions.  Specifically, we evaluate how the ratio of
the luminous mass to this dynamical mass indicator changes with radius
using the circular apertures defined in \S \ref{sec:TDG} and the light
curves given in Fig.~\ref{fig:sumIRING}. In this case, $\Delta V$ is
2.35 times the one-dimensional velocity dispersion, and $R$ is the
deconvolved aperture radius. The luminous mass is calculated from the
\hi\ mass and optical luminosity as above.

The results of this exercise is shown in Fig.~\ref{fig:MlumMdyn}.
Three curves are drawn, each one representing a different stellar
mass-to-light ratio: the solid curve shows the result for $M_* / L_B =
0$ \ML, the dotted curve shows the results for $M_* / L_B = 2$ \ML,
and the dashed line shows the results for $M_* / L_B = 5$ \ML. We see
that this ratio decreases with larger apertures. Therefore, taking a
larger radius will not change the conclusions reached above. Taking a
smaller radius might turn up objects with a larger luminous to
dynamical mass, but in this case they would have a much smaller mass
scale than is usually assumed.

In conclusion, we find no kinematic signature of a distinct dynamical
entity within the southern tail, but any signature might be masked by
the strong geometric velocity gradients in this region. If there are
bound mass concentrations, they require a significant contribution
from either evolved stars or dark matter to bind them.

If the lower distance of 13.8 Mpc suggested by Saviane \et\ (2001) is
correct, then the luminous mass decreases by a factor of 2 while the
virial mass decreases by a factor of 1.4, so the luminous to virial
mass estimates all decrease by a factor of 1.4, making it less likely
that either TDG candidate is a bound object. Additionally, if there
are bound regions, the mass scale would be more typical of dwarf
Spheroidal galaxies than dwarf Irregulars, and the gaseous
concentrations may have more relevance to remnant streams around
galaxies than the formation of dwarf irregulars (Kroupa 1998).

If there are dwarf-galaxy sized concentrations forming within the
tail, then the present observations appear inconsistent with the tidal
dwarf formation scenario suggested by Elmegreen, Kaufmann \& Thomasson
(1992). In their scenario the velocity dispersion of tidal gas is
increased over the pre-encounter value due to the interaction,
increasing the Jeans mass in the tidal gas and leading to the
formation of dwarf Irr sized objects. As noted in \S
\ref{sec:TailKin}, the velocity dispersion within the tidal tails of
NGC 4038/9 is typical of values measured in undisturbed disk
galaxies. This is true of long-tailed mergers in general (Hibbard \&
van Gorkom 1996, Hibbard \& Yun 1999a) and agrees with the expectation
from numerical models. The tails form from the dynamically coldest
material on the side of the disk opposite orbital periapses, and
expand kinematically thereafter. Tails, particularly those of
low-inclination prograde encounters, should not and appear not to be
significantly kinematically heated by the encounter.  And while it is
true that an increased velocity dispersion will suppress the formation
of low mass objects, it does not follow that it should lead to the
collapse of higher mass objects; it simply means that a higher mass is
required in order for the concentration to collapse. Therefore, the
model of tidal dwarf formation suggested by Elmegreen, Kaufmann \&
Thomasson (1992) does not appear to apply here.

\subsection{Expected Signature of Merger Remnant}

A last point that we wish to make is that NGC 4038/9 is clearly an
on-going merger experiencing a massive young starburst which is spread
throughout the disk of NGC 4038 and the overlap region. And yet, the
long tails of NGC 4038/9 suggest that the interaction began in earnest
of order 400 Myr previously. This is very similar to the situation in
the infrared luminous merger Arp 299 (Hibbard \& Yun 1999a).
Additionally, the proximity of NGC 4038/9 have allowed the
identification of a population of SSCs with ages of $\sim$ 500 Myr
(Whitmore \et\ 1999) spread throughout the disk region. Clearly, then,
merger induced extra-nuclear star-formation began around the time that
the tails were launched, has continued until today, and will likely
continue until the merger is complete.  Therefore, the merger induced
``burst" population will be broadly spread throughout the remnant,
both spatially and temporarily.  While a significant amount of the
present gas supply may find its way into the innermost regions of the
remnant, it is not clear how much of this gas will be converted into
stars. Much of this gas may be expelled in starburst driven superwind
(Heckman \et\ 1999, Hibbard \& Yun 1999b).

Either way, the expected signature of the evolved merger remnant will
not be as pronounced as is often assumed (e.g., Mihos \& Hernquist
1994, Bekki \& Shioya 1997, Silva \& Bothun 1998a,b).  Instead of a
single epoch metal-enriched population confined to the innermost radii
of the remnant and surrounded by a uniformly-old metal poor
population, there will be stars with an age range of $\sim$ 1 Gyr
spread throughout remnant body, with perhaps an inner population of
stars formed from whatever gas is not blown out.

\bigskip
\section{Conclusions}

\begin{itemize}

\item We have mapped the \hi\  in the classical double-tailed merger,
NGC 4038/9, ``The Antennae'', with unprecedented spatial and velocity
resolution. In agreement with other observers (van der Hulst 1979a,
Mahoney \et\ 1987, Gordon \et\ 2001), we find the northern tail to be
gas-rich at its tip but gas-poor at its base, the southern tail to be
gas rich along its entire length and with a significant enhancement in
the gas concentration in the vicinity of the TDG. The tail kinematics
are broadly consistent with the interaction geometry inferred from
numerical simulations. 

\item We suggest that the lack of \hi\  at the base of the northern
tail is due to photoionization by UV photons escaping the disk
starburst region. If so, the ionized gas should be readily detectable,
with and expected emission measure of order 0.8 cm$^{-6}$ pc. 

\item The \hi\ velocity field at the end of the southern tail 
is dominated by strong velocity gradients which
suggest that at this location the tail is bending away from us. 
The tail velocity gradients may mask the kinematic signature of 
any self-gravitating condensation in this region. 

\item It is not clear whether the TDG candidates identified by
Schweizer (1978) and Mirabel \et\ (1992) are self-gravitating.
The observed kinematics suggest that there is not enough mass in gas
alone to account for the \hi\ linewidths of the regions we have
delineated.  Further insight requires a measurement of the stellar
velocity dispersion and tighter constraints on the stellar
mass-to-light ratio.

\item Whether or not the TDG candidate identified by Mirabel
\et\ (1992) is a self-gravitating entity, it is clear that this
concentration of gas, stars, and star forming regions in the tail is
unique. At this location the \hi\ column densities per channel are
higher than anywhere else in the system, including within the main
disks.  We believe a clue to the origin of this concentration is that
it occurs just after the the ``V''-shape where the two filaments
connect back to each other occurs just before the location of of the
TDG.

\item These observations reveal that the ``superthin'' galaxy ESO
572--G045 is a companion to NGC 4038/9. Our low resolution data show 
a low column density \hi\  extension to the east, suggesting that this 
system may be experiencing ram pressure as it orbits NGC 4038/9. 

\end{itemize}

\acknowledgments

We thank Baerbel Koribalski for conversations and sharing of the
Gordon \et\ ATCA results prior to publication, Bill Vacca for running
population synthesis models for us, Christine Wilson and Vassilis
Charmandaris for sharing their CO map and allowing us to reproduce it
in Fig.~\ref{fig:DISKmos}f, and the referee, Pierre-Alain Duc, for a
helpful referees report.  We also thank Ed Fomalont and Juan Uson for
advice on self-calibration and mapping, Raja Guhrathakurta for help
with the optical data and reductions, Pat Smiley at NRAO for help with
the color figures, and Francois Schweizer and Jacqueline van Gorkom
for helpful discussions.  The {\sc Karma} visualization package is
highly recommended and is available at {\it
http://www.atnf.csiro.au/karma/}.


\begin{deluxetable}{llll}
\tablecaption{VLA Observing Parameters.}
\tablehead{
}\startdata
Date               &               &  1996 Jun 3 & 1997 Jun 9 \cr
Array              &               &      D      &      CnB   \cr
Time on Source (hrs) &             &     2.5     &      6.5   \cr
Phase Center ($\alpha$, $\delta$, J2000) & 
& 12$^h$ 01$^m$ 52$^s$ \hfil & --18$^\circ$ 55$'$ 40$''$\hfil \cr
Central Velocity (Heliocentric)&   &  \multispan2{~1630 \kms \hfil} \cr
Primary Beam (FWHM)&               &  \multispan2{~30$'$ \hfil}     \cr
Phase Calibrator   &               &  \multispan2{~1130--148 \hfil} \cr
Flux Calibrator    &               &  \multispan2{~3C286 \hfil}     \cr
Correlator Mode    &               &  \multispan2{~1A, on-line Hanning \hfil} \cr
Bandwidth (MHz)    &               &  \multispan2{~3.125 \hfil}     \cr
Number of Channels &               &  \multispan2{~128 \hfil}       \cr
Channel Separation (\kms) &        &  \multispan2{~5.21 \hfil}      \cr
Data set & High Resolution & Medium Resolution & Low Resolution \cr
``Robust"$^a$ parameter & -1 & +1 & +1, Convolved to 40$''$ \cr
Synthesized Beam \cr
--- Major Axis \x\  Minor Axis (FWHM) & 
$11.4'' \times 7.4''$ & $20.7'' \times 15.4''$ & $40'' \times 40''$\cr
--- Position angle (east of north) & 
76$^\circ$ & 24$^\circ$ & 0$^\circ$ \cr
Noise Level (1$\sigma$) \cr
--- Flux Density (mJy beam$^{-1}$)             & 1.3 & 0.9 & 1.3 \cr
--- Column Density (\col{19} beam$^{-1}$ \ch ) & 8.8 & 1.6 & 0.47 \cr
--- Brightness Temperature (K beam$^{-1}$)     & 9.3 & 1.7 & 0.49 \cr
\enddata
\tablenotetext{a}{Robust weighting parameter from Briggs 1995. A 
Robust parameter of -2.5 is equivalent to uniform weighting, while 
a Robust parameter of +2.5 is equivalent to Natural weighting.}
\label{tab:HIobs}
\end{deluxetable}


\begin{deluxetable}{lll}
\tablecaption{CTIO/UH Observing Parameters.}
\tablehead{
}\startdata
Date             & 1991 Apr 17/18        & 1995 Jan 22  \cr
Telescope        & CTIO 0.9m             & UH 88$''$    \cr
Camera           & Tek 512               & QUIRC       \cr
Readout Mode     & unbinned              & unbinned    \cr
Focal Ratio      & f/13.5                & f/7.5       \cr
Field of View:   &                       & \cr
--- Single CCD frame & $3\farcm8\times3\farcm8$ & $3\farcm2\times3\farcm2$ \cr
--- Final Image  & 17$'$\x17$'$          & 4$'$\x4$'$ \cr
Pixel Size       & $0\farcs439$          & $0\farcs189$ \cr
Filters          & $B,\, V,\,R$          & $K'$ (2.15 $\mu$m) \cr
Seeing           & $1\farcs8$            & $0\farcs6$  \cr
Sky Brightness (mag arcsec$^{-2}$) & 22.1, 21.3, 20.6  & --- \cr
$1\sigma$ Sky Noise, binning 1\x1 (mag arcsec$^{-2}$)
                 & 25.5, 25.3, 25.0      & --- \cr
$3\sigma$ Surface Brightness limit, binning 9\x9 (mag arcsec$^{-2}$)
                 & 27.0, 26.8, 26.5      & --- \cr
Effective Exposure Time  & $\sim$2\x600$\,$s & 3\x120$\,$s \cr
\enddata
\label{tab:OPTobs}
\end{deluxetable}


\begin{deluxetable}{lrrrrrr}
\tablecaption{Global Properties of NGC 4038/9.}
\tablehead{
\colhead{Region} &
\colhead{$\int S_{HI} dv$} &
\colhead{$M_{HI}^a$} &
\colhead{Velocity Range$^b$} &
\colhead{$\Delta\  R^c$} &
\colhead{$L_B$} &
\colhead{$M_{HI}/L_B$} \cr
&
\colhead{(Jy km s$^{-1}$)} &
\colhead{($M_\odot$)} &
\colhead{(km s$^{-1}$)} &
\colhead{(kpc)} &
\colhead{($L_\odot$)} &
\colhead{($M_\odot\  L_\odot^{-1}$)} \cr
}\startdata
Disk$^d$ &$>$17.2\+0.16 & $>1.5\times10^9$ &$>$1420--1850 &   9 & $2.4\times10^{10}$ & $>$0.06 \cr
S tail    &  32.5\+0.23 &  $2.8\times10^9$ &  1590--1770  &  65 & $3.6\times10^{9~}$ & 0.8 \cr
N tail    &   4.8\+0.10 &  $4.2\times10^8$ &  1540--1600  &  40 & $1.3\times10^{9~}$ & 0.3 \cr
Total NGC 
4038/9   &$>$54.5\+0.49 & $>4.7\times10^9$ &$>$1420--1850 & 110 & $2.9\times10^{10}$ & $>$0.2 \cr
ESO 572--G045
          &   4.3\+0.10 &  $3.7\times10^8$ &  1640--1755  &  88 & $1.9\times10^{8~}$ & 1.9 \cr
\enddata
\tablenotetext{a}{Integrated \hi\  mass, calculated using: 
$M_{HI}\,=\,2.356\times 10^5\,\Delta^2\,\int S_{HI}\,dv\, M_\odot$, 
where $\Delta$ is the distance in Mpc, and $\int S_{HI}\,dv$ is the 
integrated \hi\  emissivity, in Jy \kms. We adopt 19.2 Mpc as the 
distance to NGC 4038/9.}
\tablenotetext{b}{Range of \hi\  velocities (heliocentric), taken 
from the first-moment image.  The uncertainty is \+3 \kms.} 
\tablenotetext{c}{Maximum projected distance of the \hi\  from 
the center of NGC 4038/9.}
\tablenotetext{d}{Observations by Gordon \et\  (2001) and Gao \et\  
(2001) reveal that the cold gas within the disk and overlap
region emits over the velocity range 1340--1945 \kms. This is wider
than the bandwidth of the present observations, so that the disk and
total \hi\  fluxes derived here are lower limits.}
\label{tab:global}
\end{deluxetable}


\begin{deluxetable}{lrr}
\tablecaption{Dynamical Analysis of TDG Candidates}
\tablehead{
\colhead{Parameter}               & \colhead{TDG [MDL92]} & \colhead{TDG [S78]}\\
}\startdata
Right Ascension$^a$               & 12$^h$ 01$^m$ 25.7$^s$ \hfil  
                                  & 12$^h$ 01$^m$ 22.1$^s$ \hfil \cr
Declination$^a$                   & --19$^\circ$ 00$'$ 42$''$\hfil 
                                  & --19$^\circ$ 00$'$ 12$''$\hfil \cr
Background$^a$\\
~~~$N_{HI}$ (atoms cm$^{-2}$)     & 3.8\col{20}           & 2.9\col{20}\\
~~~$\mu_B$ (mag arcsec$^{-2}$     & 26.0                  & 26.3 \\
Half-light Radius$^a$, $r_h$
(deconvolved)                     & 3.2 kpc               & 3.5 kpc  \\
\hi\ velocity dispersion$^b$, $\sigma$  & 13.4 \kms             & 10.5 \kms \\
$M_{HI}^c$                        & 2.4\Mo{8}             & 1.7\Mo{8} \\
$L_B^c$                           & 8.7\Mo{7}             & 5.3\Mo{7} \\
$M_{vir}$                         & 1.1\Mo{9}             & 7.4\Mo{8} \\
$M_{lum}/M_{vir}\ [M_*/L_B = 0]$  & 0.3                   & 0.3 \\
$M_{lum}/M_{vir}\ [M_*/L_B = 2]$  & 0.4                   & 0.5 \\
$M_{lum}/M_{vir}\ [M_*/L_B = 5]$  & 0.7                   & 0.7 \\
\enddata        
\tablenotetext{a}{Results from fitting a Gaussian plus constant
background to the \hi\ integrated intensity map.}
\tablenotetext{b}{Result from fitting a Gaussian to an \hi\ spectrum
taken at the peak of the \hi\ distribution.}
\tablenotetext{c}{From the end-points of Fig.~\ref{fig:sumIRING}. See
text.}
\label{tab:TDG}
\end{deluxetable}
\clearpage


\begin{thebibliography}{}
\bibitem[]{}Amram, P., Marcelin, M., Boulesteix, J., \& le Coarer, E. 
	1992, A\&A, 266, 106
\bibitem[]{}Barnes, J.E. 1988, ApJ, 331, 699
\bibitem[]{}Barnes, J. E., \& Hernquist, L. 1992, Nature, 360, 715
\bibitem[]{}Bekki, K., \& Shioya, Y. 1997, ApJ, 478, L17
\bibitem[]{}Binney, J., \& Tremaine S. 1987, ``Galactic Dynamics'', 
          (Princeton: Princeton University Press). 
\bibitem[]{}Bland-Hawthorn, J., \& Maloney, P.R. 1999, ApJ, 510, 33
\bibitem[]{}Bland-Hawthorn, J., \& Putman, M. 2001, in ASP Conf.\
        Ser.\  240, Gas and Galaxy Evolution, eds.\  J.E. Hibbard, M.P. 
        Rupen \& J.H. van Gorkom (ASP, San Francisco), 369
\bibitem[]{}Bosma, A. 1981, AJ, 86, 1791
\bibitem[]{}**Braine, J., Duc, P.-A., Lisenfeld, U., Leon, S., Vallejo, O.,
     Charmandaris, V. \& Brinks, E.  2001, A\&A, in press.
\bibitem[]{}Braine, J., Lisenfeld, U., Duc, P.-A., \& Leon, S., 2000, 
     Nature, 403, 867 
\bibitem[]{}Briggs, D. 1995, PhD Thesis, NMIMT
\bibitem[]{}Burbidge, E. M., \& Burbidge, G. R. 1966, ApJ, 145, 661
\bibitem[]{}Bushouse, H. A. 1987, ApJ, 320, 49 
\bibitem[]{}Bushouse, H. A., Lamb, S. A., \& Werner, M. W. 1988, ApJ, 335, 74
\bibitem[]{}Dove, J. B. \& Shull, J. M. \& Ferrara, A. 2000, ApJ, 531, 846
\bibitem[]{}Donahue, M., Aldering, G., \& Stocke, J. T. 1995, ApJ, 450, L45
\bibitem[]{}Duc, P. -A., Mirabel, I. F. 1994, A\&A, 289, 83
\bibitem[]{}Duc, P. -A., Brinks, E., Wink, J. E., \& Mirabel, I. F. 1997, 
            A\&A, 326, 537
\bibitem[]{}Duc, P. -A., Brinks, E., Springel, V., Pichardo, B., Weilbacher, 
            P., \& Mirabel, I.F., 2000, AJ, 120, 1238 
\bibitem[]{}Elmegreen, B., Kaufmann, M., \& Thomasson, M. 1993, ApJ, 412, 90
\bibitem[]{}Evans, R., Harper, A., \& Helou, G. 1997, in Extragalactic 
         Astronomy in the Infrared, eds.\  G.A.\  Mamon, T.X.\  Thuan \& J.T. 
         Thanh Van. (Paris: Ed. Frontieres), 143
\bibitem[]{}***Evans, R. \et\ 2001, in preparation
\bibitem[]{}Fabbiano, G., Schweizer, F., \& Mackie, G. 1997, ApJ, 478, 542
\bibitem[]{}Fabbiano, G., Zezas, A., \& Murray, S.S. 2001, ApJ, 554, 1035
\bibitem[]{}Fischer, J. \et\  1996, A\&A, 315, L97
\bibitem[]{}Fritze-v.Alvensleben, U. 1998, A\&A, 336, 83
\bibitem[]{}Gao, Y., Lo, K. Y.,  Lee, S.-W., \& Lee, T.-H. 2001, ApJ, 548, 172
\bibitem[]{}Goad, J. W., \& Roberts, M. S. 1981, ApJ, 250, 79
\bibitem[]{}Gooch, R.E., 1995, in ASP Conf.~Series vol.~101, Astronomical 
    Data Analysis Software and Systems V, eds G.H.~Jacoby \& J.~Barnes 
    (ASP, SF), 80
\bibitem[]{}Gordon, S., Koribalski, B., \& Jones, K. 2001, MNRAS, 326, 578
\bibitem[]{}Graham, J. A. 1982, PASP, 94, 244
\bibitem[]{}Haas, M., Klaas, U., Coulson, I., Thommes, E., \& Xu, C. 2000, 
         A\&A, 356, L83
\bibitem[]{}Heckman, T. M., Armus, L., Weaver, K. \& Wang, J. 1999, ApJ, 
         517, 130
\bibitem[]{}Hernquist, L., \& Spergel, D. N. 1992, ApJ, 399, L117
\bibitem[]{}Hibbard, J. E., \& van Gorkom, J. H. 1996, AJ, 111, 655
\bibitem[]{}Hibbard, J. E., \& van Gorkom, J. H. 1993, 
	in ASP conf.~series, Vol.~48, The Globular Cluster--Galaxy Connection, 
	eds.\  G. H.~Smith and J. P.~Brodie (ASP, San Francisco), 619
\bibitem[]{}Hibbard, J. E., Guhathakurta, P., van Gorkom, J. H., \& 
	Schweizer, F. 1994, AJ, 107, 67
\bibitem[]{}Hibbard, J.E., \& Mihos, J.C. 1995, AJ, 110, 140.
\bibitem[]{}Hibbard, J. E., Vacca, W. D., \& Yun, M. S. 2000, AJ, 
        119, 1130. 
\bibitem[]{}Hibbard, J. E., \& Yun, M. S. 1999a, AJ, 118, 162 
\bibitem[]{}Hibbard, J. E., \& Yun, M. S. 1999b, ApJ, 522, L93
\bibitem[]{}Hoopes, C., Walterbos, R. \& Rand, R. 1999, ApJ, 522, 669
\bibitem[]{}Huchtmeier, W. K., \& Bohnenstengel, H.-D. 1975, A\&A, 41, 477
\bibitem[]{}Hummel, E., \& van der Hulst, J. M. 1986, A\&A, 155, 151
\bibitem[]{}Hunsberger, S., Charlton, J., \& Zaritsky, D. 1996, ApJ, 462, 50
\bibitem[]{}Hunsberger, S., Charlton, J., \& Zaritsky, D. 1998, ApJ, 505, 536
\bibitem[]{}Hurwitz, M., Jelinsky, P., \& Dixon, W. 1997, ApJ, 481, L31
\bibitem[]{}Iglesias-P\' aramo, J., \& V\' ilchez, J. M. 2001, ApJ, 550, 204
\bibitem[]{}Karachentsev, I. D.,  Karachentseva, V. E., \& Parnovskij, 
S. L. 1993, {\it Astronomische Nachrichten}, 314, 97
\bibitem[]{}Kaufman, M., Brinks, E., Elmegreen, D. M., Thomasson, M., 
	Elmegreen, B. G., Struck, C. \& Klaric, M.~1997, AJ, 114, 2323
\bibitem[]{}Kennicutt, R. C.~Jr. 1989. ApJ, 344, 685
\bibitem[]{}Kennicutt, R. C.~Jr., Keel, W. C., van der Hulst, J. M., 
	Hummel, E., Roettiger, K. A., 1987, AJ, 93, 1011
\bibitem[]{}**Knierman, K. A., Gallagher, S. C., Charlton, J. C., 
        Hunsberger, S. D., Whitmore, B., Kundu, A., Hibbard, J. E., \& 
        Zaritsky, D. 2001, AJ, in press
\bibitem[]{}Kroupa, P. 1998, MNRAS, 300, 200
\bibitem[]{}Kunze, D. \et\  1996, A\&A, 315, L101
\bibitem[]{}Laurent, O., Mirabel, I.F., Charmandaris, V., Gallais,
    P., Madden, S.C., Sauvage, M., Vigroux, L., \& Cesarsky, C. 2000,
    A\&A, 359, 887
\bibitem[]{}**Lipari, S., Sanders, D., Terlevich, R., Veilleux, S.,
      Diaz, R., Taniguchi, Y., Zheng, W., Kim, D., Tsvetanov, Z., 
      Carranza, G., \& Dottori, H. 2001, ApJ, submitted (astro-ph/0007316)
\bibitem[]{}Mahoney, J. M., Burke, B. F., van der Hulst, 
	J. M. 1987, in IAU Symp.~No.~117, Dark Matter in the Universe, 
	eds. ~J.~Kormendy, G.R.~Knapp (Reidel, Dordrecht), 94
\bibitem[]{}Matthews, L. D., Gallagher, J. S. III, \& van Driel,
	W. 1999, AJ, 118, 2751 (superthins)
\bibitem[]{}Mengel, S., Lehnert, M. D., Thatte, N., Tacconi-Garman,
        L. E., \& Genzel, R. 2001, ApJ, 550, 280
\bibitem[]{}Mihos, J. C. 2001, ApJ, 550, 94
\bibitem[]{}Mihos, J. C., Bothun, G. D., Richstone, D. O. 1993, ApJ, 418, 82
\bibitem[]{}Mihos, J. C., Hernquist, L. 1994, ApJ, 437, L47
\bibitem[]{}Mirabel, I. F., Dottori, H., \& Lutz, D. 1992, A\&A, 256, L19
\bibitem[]{}Mirabel, I. F., Vigroux, L., Charmandaris, V., Sauvage, M., Gallais, 
     P., Tran, D., Cesarsky, C., Madden, S. C., \& Duc, P.-A. 1998, A\&A, 333, L1
\bibitem[]{}Moore, B., \& Davis, M.~1994, MNRAS, 270, 209
\bibitem[]{}Neff \& Ulvestad 2000, AJ, 120, 670
\bibitem[]{}Nikola, T., Genzel, R., Herrmann, F., Madden, S.C., Poglitsch, A.,
   Geis, N., Townes, C.H., \& Stacey, G.J. 1998, ApJ, 504, 749
\bibitem[]{}Peterson, S. D., \& Shostak, G. S. 1974, AJ, 79, 767
\bibitem[]{}Read, A. M., Ponman, T. J., \& Wolstencroft 1995, MNRAS, 277, 397
\bibitem[]{}Rubin, V. C., Ford, W. K. Jr, \& D'Odorico, S. 1970, ApJ, 160, 801
\bibitem[]{}Rupen, M. P. 1999, in {\it Synthesis Imaging in Radio Astronomy}, 
        ASP Conf. Series vol. 180, ed. G. B. Taylor,
	C. L. Carilli and R. A. Perley (ASP, San Francisco), 229
\bibitem[]{}Sansom, A. E., Dotani, T., Okada, K., Yamashita, A., \& Fabbiano,
             G. 1996, MNRAS, 281, 48 
\bibitem[]{}***Saviane, I., Rich, R.M.R., \& Hibbard, J.E. 2001, in preparation
\bibitem[]{}Schombert, J. M., Wallin, J. F., \& Struck-Marcell, C. 1990, AJ, 99, 497
\bibitem[]{}Schweizer, F. 1978, The Structure and Properties of 
    Nearby Galaxies, IAU Symp.~No.~77, edited by~E. M.~Berkhuijsen and 
    R.~Wielebinski (Reidel, Dordrecht), 279
\bibitem[]{}Schweizer, F. 1998, in Saas-Fee Advanced Course No.~26, 
        Galaxies: Interactions 
	and Induced Star Formation, R. C.~Kennicutt Jr.,
	F.~Schweizer and J.~E.~Barnes (Springer, Berlin), 105
\bibitem[]{}Silva, D. R., \& Bothun, G. D. 1998a, AJ, 116, 85
\bibitem[]{}Silva, D. R., \& Bothun, G. D. 1998b, AJ, 116, 2793
\bibitem[]{}Smith, B. J., \& Higdon, J. L. 1994, AJ, 108, 837
\bibitem[]{}Smith, B. J., \& Struck, C. 2001, ApJ, 121, 710
\bibitem[]{}Stanford, S. A., Sargent, A. I., Sanders, D. B., \& 
	Scoville, N. Z. 1990, ApJ, 349, 492
\bibitem[]{}Stewart, S. G., Fanelli, M. N., Byrd, G. G., Hill, J. K., 
  Westpfahl, D. J., Cheng, K.-P., O'Connell, R. W., Roberts, M. S., Neff, 
  S. G., Smith, A. M., \& Stecher, T. P. 2000, ApJ, 529, 201
\bibitem[]{}Taff, L. G., Lattanzi, M. G., Bucciarelli, B., Gilmozzi,
	R., McLean, B. J., Jenkner, H., Laidler, V. G., Lasker, B. M., Shara,
	M. M., \& Sturch, C. R. 1990, ApJ, 353, L45
\bibitem[]{}Tenorio-Tagle, G., Bodenheimer, P. 1988, ARA\&A, 26, 145
\bibitem[]{}Toomre, A., \& Toomre, J. 1972, ApJ, 178, 623
\bibitem[]{}van der Hulst, J. M. 1979a, A\&A, 155, 151
\bibitem[]{}van der Hulst, J. M. 1979b, A\&A, 75, 97
\bibitem[]{}van der Kruit, P. C. 1988, A\&A, 192, 117
\bibitem[]{}Vigroux, L. \et\  1996, A\&A, 314, L93 
\bibitem[]{}Wainscoat, R. J., \& Cowie, L. L. 1992, AJ, 103, 332
\bibitem[]{}Walter, F., \& Brinks, E. 1999, AJ, 118, 273
\bibitem[]{}Weilbacher, P. M., Duc, P.-A., Fritze-v.Alvensleben, U., Martin, P.,
        \& Fricke, K. J. 2000, A\&A, 358, 819
\bibitem[]{}Whitmore, B. C., \& Schweizer, F. 1995, AJ, 109, 960
\bibitem[]{}Whitmore, B. C., Zhang, Q., Leitherer, C., Fall, S. M., 
	Schweizer, F., Miller, B. W., 1999, AJ, 118, 1551
\bibitem[]{}Wilson, C. D., Scoville, N., Madden, S. C., \& Charmandaris, 
         V. 2000, ApJ, 542, 120
\bibitem[]{}Xu, C., Gao, Y., Mazzarella, J., Lu, N., Sulentic, J. W., \& 
         Domingue, D.L. 2000, ApJ, 546, 644
\bibitem[]{}Yang, H., Chu, Y.-H., Skillman, E. D., \& Terlevich, R., 1996, 
     AJ, 112, 146 
\bibitem[]{}Yun, M. S. 1999, in IAU Symp. No.~186, Galaxy Interactions at Low and 
        High Redshift, ed. D. Sanders \& J. Barnes (City: Publisher), 81 
\bibitem[]{}Yun, M. S., Ho, P. T. P., \& Lo, K. Y. 1994, Nature, 372, 530
\bibitem[]{}**Zhang, Q., Fall, S. M., \& Whitmore 2001, ApJ, in press 
          (astro-ph/0105174)
\bibitem[]{}Zhu, M. 2001, Ph.D. Thesis, University of Toronto
\bibitem[]{}Zwicky, F. 1956, {\it Ergebnisse der Exakten Naturwissenschaften}, 
      29, 344
\end{thebibliography}
\end{document}